\documentclass[english,a4paper,11pt]{article}
\pdfoutput=1
\usepackage{jheppub}

\usepackage[T1]{fontenc}
\usepackage[latin9]{inputenc}
\usepackage{verbatim}
\usepackage{amstext}
\usepackage{amssymb}
\usepackage{graphicx}

\makeatletter
\def\@fpheader{\relax}

\providecommand{\tabularnewline}{\\}

\@ifundefined{textcolor}{}
{%
 \definecolor{BLACK}{gray}{0}
 \definecolor{WHITE}{gray}{1}
 \definecolor{RED}{rgb}{1,0,0}
 \definecolor{GREEN}{rgb}{0,1,0}
 \definecolor{BLUE}{rgb}{0,0,1}
 \definecolor{CYAN}{cmyk}{1,0,0,0}
 \definecolor{MAGENTA}{cmyk}{0,1,0,0}
 \definecolor{YELLOW}{cmyk}{0,0,1,0}
}

\makeatother

\usepackage{babel}
\begin{document}

\title{Why Is The Neutrino Oscillation Formula Expanded In $\Delta m_{21}^{2}/\Delta m_{31}^{2}$
Still Accurate Near The Solar Resonance In Matter?}

\author{Xun-Jie Xu}

\emailAdd{xunjie.xu@gmail.com}

\selectlanguage{english}%

\affiliation{Max-Planck-Institut f\"{u}r Kernphysik, Postfach 103980, D-69029
Heidelberg, Germany.\\
Institute of Modern Physics and Center for High Energy Physics, Tsinghua
University, Beijing 100084, China.}
\abstract{
The conventional approximate formula for neutrino oscillation in matter
which is obtained from the expansion in terms of the ratio of mass
square differences $\alpha=\Delta m_{21}^{2}/\Delta m_{31}^{2}\approx0.03$,
first proposed by Cervera, et al and Freund, turns out to be an accurate
formula for accelerator neutrino experiments. Originally it required
the neutrino energy to be well above the solar resonance to validate
the expansion but it is found to be still very accurate when the formula
is extrapolated to the resonance, which is practically important for
the T2K experiment. This paper shows that the accuracy is guaranteed
by cancellations of branch cut singularities and also, for the first
time, analytically computes the actual error of the formula. The actual
error implies that the original requirement can be safely removed
in current experiments.  
}
\maketitle

\section{Introduction}

In long-baseline(LBL) neutrino experiments, the matter effect\cite{Wolfenstein:1977ue,Mikheev:1986gs,Mikheev:1986wj}
is usually not negligible. For current LBL accelerator neutrino experiments
such as T2K\cite{T2K2013PRL,T2K2014PRL}, MINOS\cite{MINOS} and NOvA\cite{NOVA,NOVA2}
where the matter densities are almost constant, there is a useful
approximate formula for the transition probability. Taking the same notations as PDG,
 the formula is\cite{PDG2014}
\begin{eqnarray}
P(\nu_{\mu}\rightarrow\nu_{e}) & = & 4s_{13}^{2}c_{13}^{2}s_{23}^{2}\frac{\sin^{2}(1-A)\Delta}{(1-A)^{2}}\nonumber \\
 &  & +8\alpha\frac{J_{CP}}{s_{\delta}}\cos(\Delta+\delta)\frac{\sin A\Delta}{A}\frac{\sin(1-A)\Delta}{1-A}\nonumber \\
 &  & +4\alpha^{2}s_{12}^{2}c_{12}^{2}c_{23}^{2}\frac{\sin^{2}A\Delta}{A^{2}}\label{eq:001}
\end{eqnarray}
where 
\begin{equation}
A\equiv 2\sqrt{2}G_{F}N_{e}E/\Delta m_{31}^{2},\thinspace\alpha\equiv\Delta m_{21}^{2}/\Delta m_{31}^{2}\approx0.03,\label{eq:0205-1}
\end{equation}
and $\Delta\equiv\Delta m_{31}^{2}L/(4E)$. $N_{e}$ is the electron
number density in matter, about $1.4\textrm{cm}^{-3}N_{A}$ in the
Earth's crust. 

The formula was originally derived in \cite{Cervera:2000kp,Freund:2001pn}
as a  series expansion in $\alpha$. But the problem is that due to
the non-perturbative behavior near the solar resonance, the expansion 
is expected to be valid only when the neutrino energy is well above the 
solar resonance, 
\begin{equation}
E\gg 0.34\textrm{GeV}\frac{\Delta m_{21}^{2}}{7.6\times10^{-5}\textrm{eV}^{2}}\frac{1.4\textrm{cm}^{-3}N_{A}}{N_{e}}.\label{eq:1225-3}
\end{equation}
This was emphasized in ref.\cite{Freund:2001pn}, because the approximation
$\alpha/A\ll1$ was used when the formula was derived. We will reformulate
the derivation of the formula in section \ref{sec:Accurarcy-of--expansion}
to show the problem more explicitly but here we take the solar mixing
angle $\theta_{12}$ as a good example to show the problem. The effective
$\sin2\theta_{12}$ in matter, denoted as $\sin2\theta_{12}^{m}$,
expanded in $\alpha$ to first order, is \cite{Freund:2001pn}
\begin{equation}
\sin2\theta_{12}^{m}\sim\alpha/A.\label{eq:0129}
\end{equation}
The solar resonance is at $A=\alpha\cos2\theta_{12}\approx0.4\alpha$
so near the solar resonance $\sin2\theta_{12}^{m}$ is
quite likely to be larger than 1. As will be shown in section \ref{sec:Accurarcy-of--expansion},
$\sin2\theta_{12}^{m}>1$ does appear in the expansion when the energy
is lower than $0.34\textrm{GeV}$, which makes the calculation 
invalid. Originally, $\sin2\theta_{12}^{m}$ in the calculation was
expected not only less than 1 but also small, i.e. $\sin2\theta_{12}^{m}\ll1$,
otherwise the unitarity of the effective mixing matrix will be badly
violated, thereby invalidating the calculation. 

Despite the claimed bound (\ref{eq:1225-3}) in \cite{Freund:2001pn},  in practice this formula works well below
the bound (see figures \ref{fig:proba-all}, \ref{fig:proba-all-1} presented later in this paper). For example T2K 
has used this formula in their recent publication\cite{T2K2013PRD}
because eq.(\ref{eq:001}) exhibits excellent accuracy near the solar
resonance
\footnote{Note that for T2K, the energy range is $0.1$-1.2 GeV and the spectrum
peaks at $0.6$ GeV\cite{T2K2014PRL}. A part of the current measured
range $0.1$-0.34 GeV is below the bound (\ref{eq:1225-3}) which
would lead to $\sin2\theta_{12}^{m}>1$ in the expansion.%
}. 

So  (\ref{eq:1225-3}) is most likely not the true bound of validity. 
We would like to know  to what extent  the formula is accurate
or valid. The main goal of this paper, is to mathematically demonstrate that 
there is no lower bound of $A$ for the domain of validity. 
We will provide explicit errors of the formula, among which the main error
related to the matter effect is only 
$\mathcal{O}(s_{13}^{2}\alpha A\Delta^{2})$. This implies that the formula is 
still accurate when $A$ is close to $\alpha$ and one may apply (\ref{eq:001}) below 
the bound. 


Note that a higher order calculation in the original perturbative approach
will not work since the series in $\alpha/A$  can not converge 
at the resonance if the branch cut singularity is not treated carefully.
Actually a higher order correction to the formula (\ref{eq:001})
is computed in ref.\cite{Asano:2011nj} but  the correction blows up when taking
the vacuum limit $A\rightarrow0$. Thus it can not give a correct
estimation when $A$ is small. This is due to a
lack of careful treatment of the branch cut singularity related to
the solar resonance. 

Branch cuts in the oscillation system with the matter effect are essentially related to level crossings
\cite{Akhmedov:2008qt,Blennow:2013rca}, but
less noticed before. Note that the three eigenvalues of the oscillation
system come from the same cubic equation but they are different. The
difference originates from the different branches in the square roots
and cubic roots in the general solutions of a cubic equation. At a
level crossing two of the eigenvalues are very close to each other
which makes the problem quite non-perturbative and this just corresponds
to the starting point of the branch cuts, which are called branch
cut singularities. The branch cut singularities are essentially  origins of
all non-perturbativities in the oscillation system. In this paper, we will remove
the singularity corresponding to the solar resonance in our analytic
calculation by transformation of the eigenvalues to some singularity-free
variables and compute the $S$-matrix using the Cayley-Hamilton theorem.
In this way the conventional formula will be proven to be accurate below
the bound (\ref{eq:1225-3}). The relation between the branch cut
singularities and level crossings will be discussed in detail and
thus improve our understanding of the matter effect in neutrino oscillation\cite{Akhmedov:2001kd,Akhmedov:2006hb,Xing:2013uxa,Chiu:2010da,Takamura:2005df,Arafune:1997hd,Blennow:2004js,Ohlsson:2003ip,Ohlsson:2013ip,Schwetz:2007py,Yokomakura:2002av,Smirnov:2013cqa,Freund:1999gy,Zhang:2006yq}. 

As a byproduct of our analysis, a new approximate formula is 
derived in this paper, with better accuracy. 
Though the exact form is a little more complicated than (\ref{eq:001}), 
for practical use in neutrino simulation, it is useful and covers most aspects. 
This is important considering that simulation 
of LBL experiments and performing  $\chi^{2}$-fits
 require a fast and simple method to compute a large 
 number of oscillation probabilities. 
Therefore even though the numerical calculation is always viable,
there are still many studies on analytic approximation formulae for neutrino
oscillation in matter\cite{Akhmedov:2004ny,Honda:2006hp,Agarwalla:2013tza,Zhou:2011xm,Takamura:2004cr,Takamura:2005yi,Harrison:2002ee,Kimura:2002hb,Kimura:2002wd,Ohlsson:2001vp,Yasuda:2014hwa,Asano:2011nj,Minakata:2015gra}.

This paper is organized as follows. In section \ref{sec:Accurarcy-of--expansion},
we reformulate the original derivation of the formula (\ref{eq:001}) and numerically
show the accuracy of the $\alpha$-expansion in the case of T2K.
We will see that the $\alpha$-expansion for some effective parameters
is actually invalid below $0.34\textrm{GeV}$ in T2K while the final
result of the probability is very accurate. Then in section \ref{sec:The-problem-of}
from the viewpoint of singularities, we show that non-differentiable
singularities in many parameters originate from the branch cut
and result in the failure of the $\alpha$-expansion.
In section \ref{sec:Solution} we solve the problem rigorously and
then compute the analytical error of (\ref{eq:001}). Based
on the calculation in section \ref{sec:Solution}, we also propose
an alternative to the conventional formula. 
Their accuracies are numerically verified, which will be shown in
section \ref{sec:discuss}. Finally we conclude in section \ref{sec:Conclusion}.

\section{\label{sec:Accurarcy-of--expansion}The $\alpha$-expansion and the
accidental accuracy }

In this section, we first introduce analytic diagonalization of the
$3\times3$ effective Hamiltonian, which has early been done by Zaglauer
and Schwarzer \cite{Zaglauer:1988gz} without any approximation. Then
we show the $\alpha$-expansion of the result from Freund's
calculation\cite{Freund:2001pn} and compare the approximate result
with the exact result (though complicated but numerically programmable) to
see how much it deviates from the exact result. We will show that
the $\alpha$-expansion result of effective neutrino parameters are
quite inaccurate and even invalid near the solar resonance but the
final result (i.e. the assembled oscillation probability)
from these parameters is very accurate.

Neutrino oscillation in matter is subjected to the Schr\"{o}dinger
equation in the flavor space, 
\begin{equation}
i\frac{d}{dL}|\nu(L)\rangle=H|\nu(L)\rangle,\label{eq:1228}
\end{equation}
where $|\nu(L)\rangle$ denotes the flavor state of the evolving neutrino
at a distance of $L$ from the source and $H$ is the Hamiltonian
represented by the $3\times3$ matrix in the standard neutrino oscillation framework,
\begin{equation}
H=\frac{1}{2E}U.\left(\begin{array}{ccc}
m_{1}^{2}\\
 & m_{2}^{2}\\
 &  & m_{3}^{2}
\end{array}\right).U^{\dagger}+\sqrt{2}G_{F}N_{e}\left(\begin{array}{ccc}
1\\
 & 0\\
 &  & 0
\end{array}\right).\label{eq:1227-1}
\end{equation}
Here $U$ and $m_{i}$'s  are neutrino mixing matrix and  masses in vacuum respectively. The second
 term in eq.(\ref{eq:1227-1}) comes from the matter effect.
Without the second term (i.e. $N_{e}=0$), the solution of (\ref{eq:1228}) is quite
simple since the first term has
already been diagonalized. So in vacuum, the transition amplitude  of $\nu_{\alpha}\rightarrow\nu_{\beta}$
is just
\begin{equation}
S_{\alpha\beta}=U_{\alpha1}U_{\beta1}^{*}+U_{\alpha2}U_{\beta2}^{*}e^{i\alpha\Delta}+U_{\alpha3}U_{\beta3}^{*}e^{i\Delta}.\label{eq:1228-1}
\end{equation}
Here $S_{\alpha\beta}$ is usually referred to as the $S$-matrix in neutrino
oscillation. 
For neutrino oscillation in matter, we need to diagonalize
(\ref{eq:1227-1}) to obtain the effective mixing matrix $\tilde{U}$
and the effective neutrino masses $\tilde{m}_{k}$, defined as 
\begin{equation}
H=\frac{1}{2E}\tilde{U}diag(\tilde{m}_{1}^{2},\tilde{m}_{2}^{2},\tilde{m}_{3}^{2})\tilde{U}^{\dagger}.\label{eq:1229}
\end{equation}
Then $\tilde{U}$ and $\tilde{m}_{k}$, combined in the way similar to (\ref{eq:1228-1}), gives  the $S$-matrix in
matter. 

The $3\times3$ matrix $H$ can be analytically diagonalized by 
solving first the eigenvalues and then 
the corresponding eigenvectors, though the computation is complicated.

The process can be a little simplified if we extract a
dimensionless matrix $M$ from
\begin{equation}
H=\frac{m_{1}^{2}}{2E}+\frac{\Delta m_{31}^{2}}{2E}M,\label{eq:1228-2}
\end{equation}
and define
\begin{equation}
M_{d}=U^{\dagger}MU.\label{eq:0107-1}
\end{equation}
Then $M_{d}$ is 
\begin{equation}
M_{d}=\left(\begin{array}{ccc}
0\\
 & \alpha\\
 &  & 1
\end{array}\right)+Au^{T}.u,\label{eq:1023-13}
\end{equation}
where $u\equiv(U_{e1},U_{e2},U_{e3})$ is the first row of $U$ and
is real by proper rephasing $U$. The cubic equation for the
eigenvalues of $M_{d}$ is
\begin{equation}
\lambda^{3}+b\lambda^{2}+c\lambda+d=0,\label{eq:1228-4}
\end{equation}
with
\begin{equation}
b=-1-A-\alpha;\thinspace c=A-Au_{3}^{2}+\alpha+A\alpha-Au_{2}^{2}\alpha;\thinspace d=-A\alpha u_{1}^{2}.\label{eq:1228-5}
\end{equation}
The eigenvalues of $M_{d}$ (Note that $M$   has the same
eigenvalues as $M_{d}$) solved from eq.(\ref{eq:1228-4}) are
\begin{equation}
\lambda_{k+1}=-\frac{1}{3}(b+e^{-2k\pi i/3}\Delta_{3}+e^{2k\pi i/3}\overline{\Delta_{3}}),\label{eq:1228-3}
\end{equation}
where $k=0,1,2$ and 
\begin{equation}
\Delta_{0}=b^{2}-3c;\Delta_{1}=2b^{3}-9bc+27d;\Delta_{3}=\left(\frac{\Delta_{1}+i\sqrt{4\Delta_{0}^{3}-\Delta_{1}^{2}}}{2}\right)^{\frac{1}{3}}.\label{eq:1228-6}
\end{equation}
Then the effective neutrino masses defined in eq.(\ref{eq:1229})
are given by
\begin{equation}
\tilde{m}_{k}^{2}=m_{1}^{2}+\Delta m_{31}^{2}\lambda_{k},\label{eq:1229-1}
\end{equation}
which can be expressed in terms of $\alpha$ and $A$ explicitly  according
to eqs.(\ref{eq:1228-5}), (\ref{eq:1228-3}) and (\ref{eq:1228-6}). 

Then $\tilde{U}$ can be computed by solving the corresponding eigenvectors
of $\lambda_{k}$. 
The reader may refer to \cite{Zaglauer:1988gz}
for the full form of eigenvectors. Once $\tilde{U}$ is computed, 
we can extract effective mixing angles from the standard parametrization of $\tilde{U}$.
All  effective parameters (masses and mixing angles)
expanded in  $\alpha$ are given below\cite{Freund:2001pn}: 
\begin{eqnarray}
\lambda_{1} & = & \frac{1}{2}(1+A-C)+\frac{\alpha(C+1-A\cos2\theta_{13})s_{12}^{2}}{2C}+\mathcal{O}(\alpha^{2}),\label{eq:1231}\\
\lambda_{2} & = & \alpha c_{12}^{2}+\mathcal{O}(\alpha^{2}),\label{eq:1231-1}\\
\lambda_{3} & = & \frac{1}{2}(1+A+C)+\frac{\alpha(C-1+A\cos2\theta_{13})s_{12}^{2}}{2C}+\mathcal{O}(\alpha^{2}),\label{eq:1231-2}
\end{eqnarray}
where
\begin{equation}
C=\sqrt{(1-A)^{2}+4As_{13}^{2}}.\label{eq:1231-3}
\end{equation}
The effective mixing angles in matter are (we use a superscript $^{m}$
to distinguish them from vacuum parameters) 
\begin{eqnarray}
\sin^{2}2\theta_{13}^{m} & \approx & \frac{\sin^{2}2\theta_{13}}{C^{2}}+\alpha\frac{2A(\cos2\theta_{13}-A)s_{12}^{2}\sin^{2}2\theta_{13}}{C^{4}},\label{eq:1231-4}\\
\sin2\theta_{12}^{m} & \approx & \frac{\sqrt{2}\alpha C\sin2\theta_{12}}{Ac_{13}\sqrt{C(-A+C+\cos2\theta_{13})}},\label{eq:1231-5}\\
\sin2\theta_{23}^{m} & \approx & \sin2\theta_{23}+\alpha\cos\delta\frac{2A\sin2\theta_{12}s_{13}\cos2\theta_{23}}{1+C-A\cos2\theta_{13}},\label{eq:1231-6}\\
\sin\delta^{m} & \approx & \sin\delta(1-\alpha\frac{\cos\delta}{\tan2\theta_{23}}\frac{2A\sin2\theta_{12}s_{13}}{1+C-A\cos2\theta_{13}}).\label{eq:1231-7}
\end{eqnarray}

In vacuum it
is straightforward to get the expansion
\begin{equation}
P^{\textrm{vac}}=4s_{13}^{2}c_{13}^{2}s_{23}^{2}\sin^{2}\Delta+8\frac{J_{CP}}{s_{\delta}}\alpha\Delta\cos(\Delta+\delta)\sin\Delta+4s_{12}^{2}c_{12}^{2}c_{23}^{2}\alpha^{2}\Delta^{2}.\label{eq:0517}
\end{equation}
So one can replace the corresponding vacuum parameters in (\ref{eq:0517})
with the effective parameters in matter given above. This will produce  
the conventional formula (\ref{eq:001}).


\begin{figure}
\centering\includegraphics[width=8.1cm]{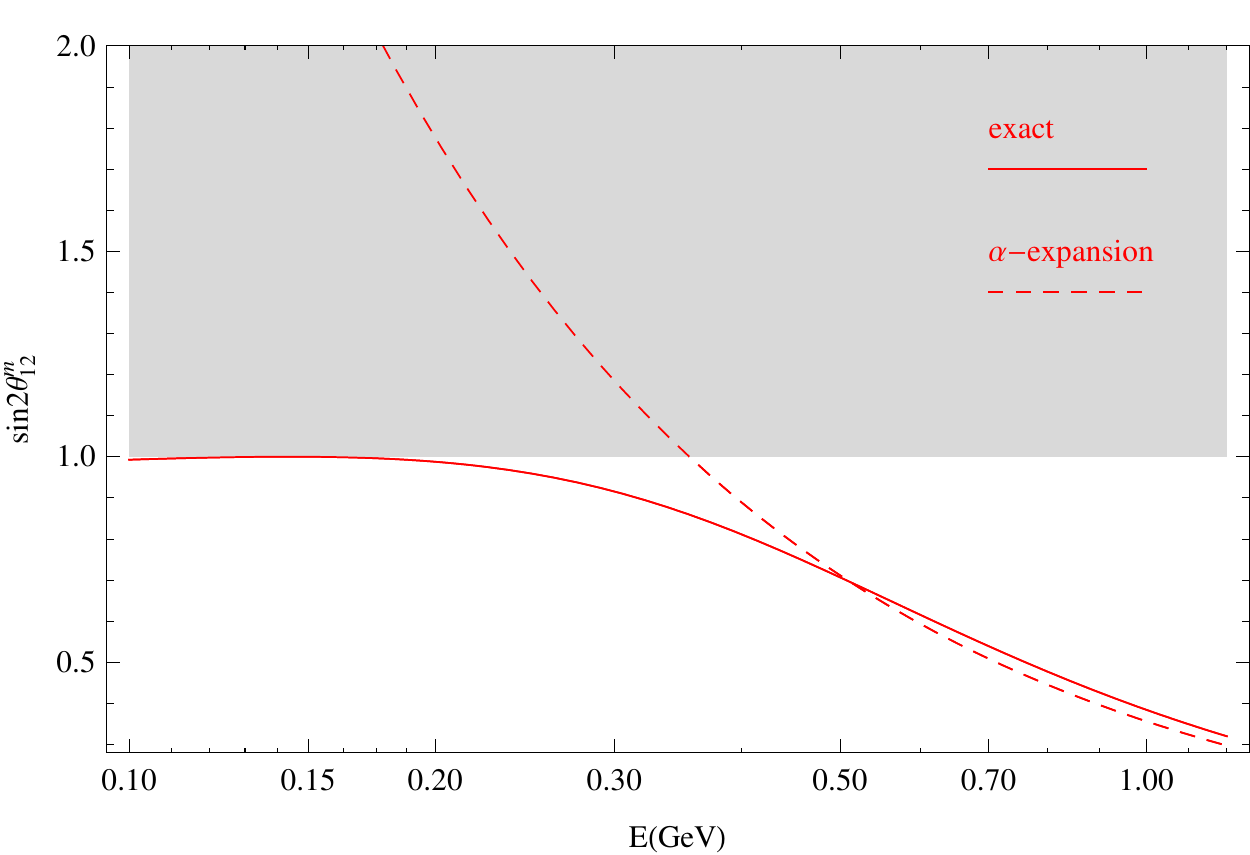}

\protect\caption{\label{fig:mixing} Compare the approximate formula of $\sin2\theta_{12}^{m}$
with the exact solution. This figure shows the invalidity of the $\alpha$-expansion
of $\sin2\theta_{12}^{m}$ in the case of T2K. When $E<0.5$GeV, it becomes inaccurate
and for $E<0.35$ GeV the result is completely invalid since the sine
value should not be larger than $1$(the gray region).}
\end{figure}

Note that the above $\alpha$-expansion of effective parameters requires
not only $\alpha\ll1$ but also $\alpha/A\ll1$. 
If we look at the effective mixing angles in (\ref{eq:1231-4}-\ref{eq:1231-7}), 
we  find that the
$\alpha$-expansion of $\sin2\theta_{12}^{m}$ may have a problem at
$\alpha/A\sim1$. In (\ref{eq:1231-5}) we see $\sin2\theta_{12}^{m}\sim\alpha/A$
which implies the correction from $\alpha$ is amplified by $1/A$,
so the expansion may be not accurate if $A$ is small. We compare it with the exact value
from \cite{Zaglauer:1988gz} in figure \ref{fig:mixing}, where the
energy range is $0.1-1.2$GeV and matter density 
is 1.3$\textrm{g/}\textrm{cm}^{3}$ (the case of the T2K experiment). 
From figure \ref{fig:mixing} we see the expansion formula fits the
exact solution well only at $E>0.5$ GeV but deviates from it quickly
when $E<0.5$GeV. More seriously, when the energy goes below 0.35
GeV then $\sin2\theta_{12}^{m}$ will be larger than $1$ (the gray region). This is because the unitarity
of $\tilde{U}$ is badly violated. 

\begin{figure}
\centering\includegraphics[width=8.1cm]{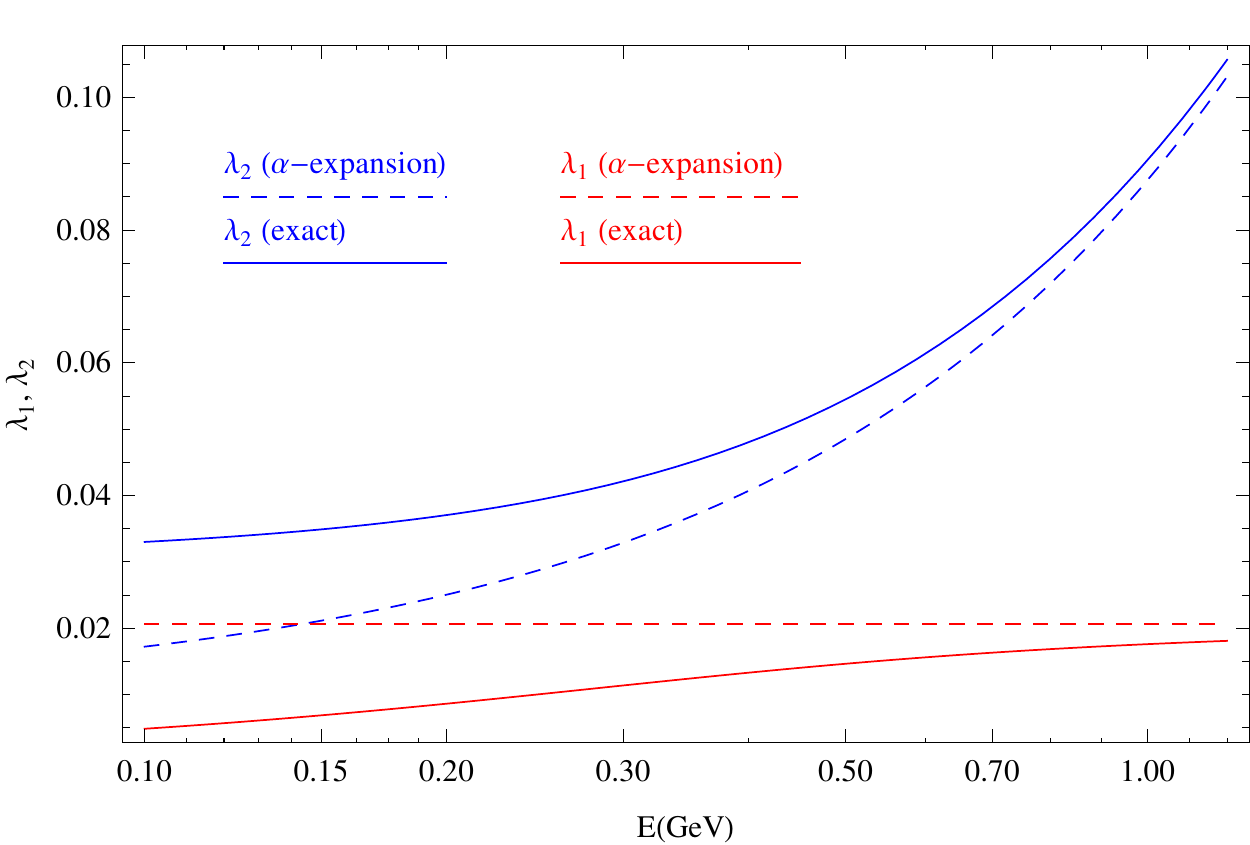}

\protect\caption{\label{fig:mass}Compare the approximate formula of the eigenvalues
$\lambda_{1}$ and $\lambda_{2}$ with the exact solution.}
\end{figure}

The other parameters suffering from the same problem are $\lambda_{1}$
and $\lambda_{2}$. We plot them with the exact solutions in figure
\ref{fig:mass}. We see that below 0.3 GeV the effective mass square difference $\Delta\tilde{m}_{21}^{2}=m_{3}^{2}(\lambda_{2}-\lambda_{1})$
from the exact solution (solid curves) can be several times that of  the $\alpha$-expansion (dashed curves), which also implies 
invalidity of the $\alpha$-expansion at low energies.

\begin{figure}
\centering\includegraphics[width=7.5cm]{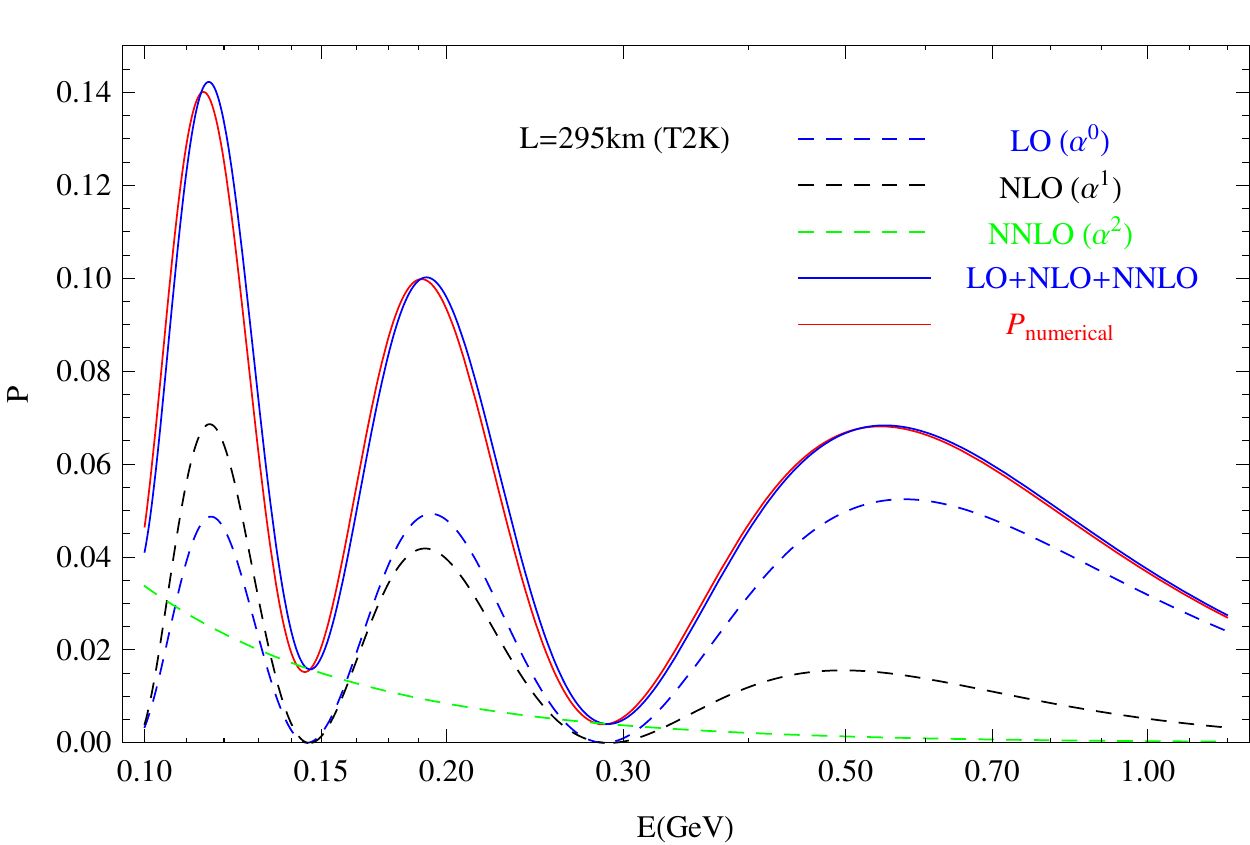}%

\protect\caption{\label{fig:comp1}The plot shows that the conventional formula (blue solid curve) given by (\ref{eq:001}) is very accurate when used in T2K, in contrast with
figure \ref{fig:mixing} and figure \ref{fig:mass} where those effective
parameters used to compute the probability are very inaccurate. All 
three order contributions ($\alpha^0$, $\alpha^1$ and $\alpha^2$)
 are also plotted (dashed curves) to show that all of them are  indispensable to make (\ref{eq:001}) accurate in T2K. }
\end{figure}

But interestingly, the formula of oscillation probability assembled
from these inaccurate and even invalid pieces is very accurate, as
shown in figure \ref{fig:comp1} where we use the
same energy range and matter density as figure \ref{fig:mixing} and
figure \ref{fig:mass}. 

One argument might be that, the accuracy of $P$ is because $\sin2\theta_{12}^{m}$
does not appear at the leading order (LO) of (\ref{eq:001}), but only at the next-to-leading order (NLO) and the next-to-next-to-leading order (NNLO) which are of order $\alpha^1$ and $\alpha^2$, respectively. 
This  suppresses the effect of the inaccuracy from
$\sin2\theta_{12}^{m}$ shown in figure \ref{fig:mixing}. But in
figure \ref{fig:comp1} we see that at the second and third peaks
(from right to left), the NLO and NNLO are comparable to the LO so
the accuracy can not be explained by the NLO suppression. 

It might be expected that the calculation at a higher order of $\alpha$
can explain this by finding some cancellations between errors. However,
at a higher order, the accuracy in figures \ref{fig:mixing},\ref{fig:mass}
turns out to be improved very little. Actually, as will be discussed
in the next section, there is an underlying problem 
that some variables in the calculation are not differentiable at $\alpha=A=0$.
For these variables, the expansion series including $\alpha/A$ can
not even converge if $\alpha/A\gtrsim1$. That is why a higher order
calculation can not solve the accuracy problem.

\section{\label{sec:The-problem-of}Non-differentiabilities, singularities
and branch cuts in the oscillating system }

To reveal the key problem in the expansion, we start from 
an oversimplified but heuristic problem, series
expansion of the following function 
\begin{equation}
g(\alpha,A)=\sqrt{\alpha^{2}+A^{2}}.\label{eq:0402}
\end{equation}
If $\alpha$ is small, but $A$ is relatively not, then we can  expand it in $\alpha$, 
\begin{equation}
g(\alpha,A)=A+\frac{\alpha^{2}}{2A}+\alpha\mathcal{O}(\frac{\alpha^{3}}{A^{3}}).\label{eq:0402-1}
\end{equation}
Here we see $\alpha/A\ll1$ is necessary to make eq.(\ref{eq:0402-1})
accurate. If $A$ is much smaller than $\alpha$, then we can expand
it in $A$ as $g(\alpha,A)=\alpha+A^{2}/(2\alpha)+\cdots$. But
what if $A$ is close to $\alpha$?  One may think that if A is close
to $\alpha$, then both are small so we can make a double expansion
of the function, 
\begin{equation}
g(\alpha,A)=c_{0}+c_{1}\alpha+c_{2}A+\mathcal{O}(\alpha^{2},\alpha A,A^{2}),\label{eq:0410}
\end{equation}
where $c_{0}=g(0,0)$, $c_{1}$ and $c_{2}$ are the partial derivatives
$\partial_{\alpha}g$ and $\partial_{A}g$ at $\alpha=A=0$. However, we cannot expand 
$\sqrt{\alpha^{2}+A^{2}}$ in this way because
the partial derivatives $c_{1}$ and $c_{2}$ do not exist (as one
can check explicitly). Geometrically this is  easy to understand
since $g=\sqrt{\alpha^{2}+A^{2}}$ is a cone in the $\alpha-A-g$
space. Expansion at the tip of the cone will certainly fail. 

Actually the function (\ref{eq:0402}) does exist in the eigenvalues
of the oscillation system%
\footnote{The eigenvalues (and thus the corresponding oscillation parameters)
contain more complicated square roots like $\sqrt{\alpha^{2}+A^{2}c_{13}^{2}-2\alpha A\kappa}$
where $\kappa\simeq1/3$ (see, e.g., calculation in \cite{Agarwalla:2013tza}),
but the problem caused by $\alpha\sim A$ in the expansion is essentially
the same as the simplified one in (\ref{eq:0402}).%
} so in the original derivation of  (\ref{eq:001}) $\alpha/A\ll1$
has to be assumed. If we use formulae derived from such an expansion
but simply ignore the condition $\alpha/A\ll1$, then we are at the
risk of getting total wrong results, such as $\sin2\theta_{12}^{m}$>1
shown in Fig.\ref{fig:mixing}. 

So  basically the question is why for the oscillation probablity
this problematic expansion works  very well. This will be answered
next by branch cut singularities.

\begin{figure}
\centering

\includegraphics[width=7.5cm]{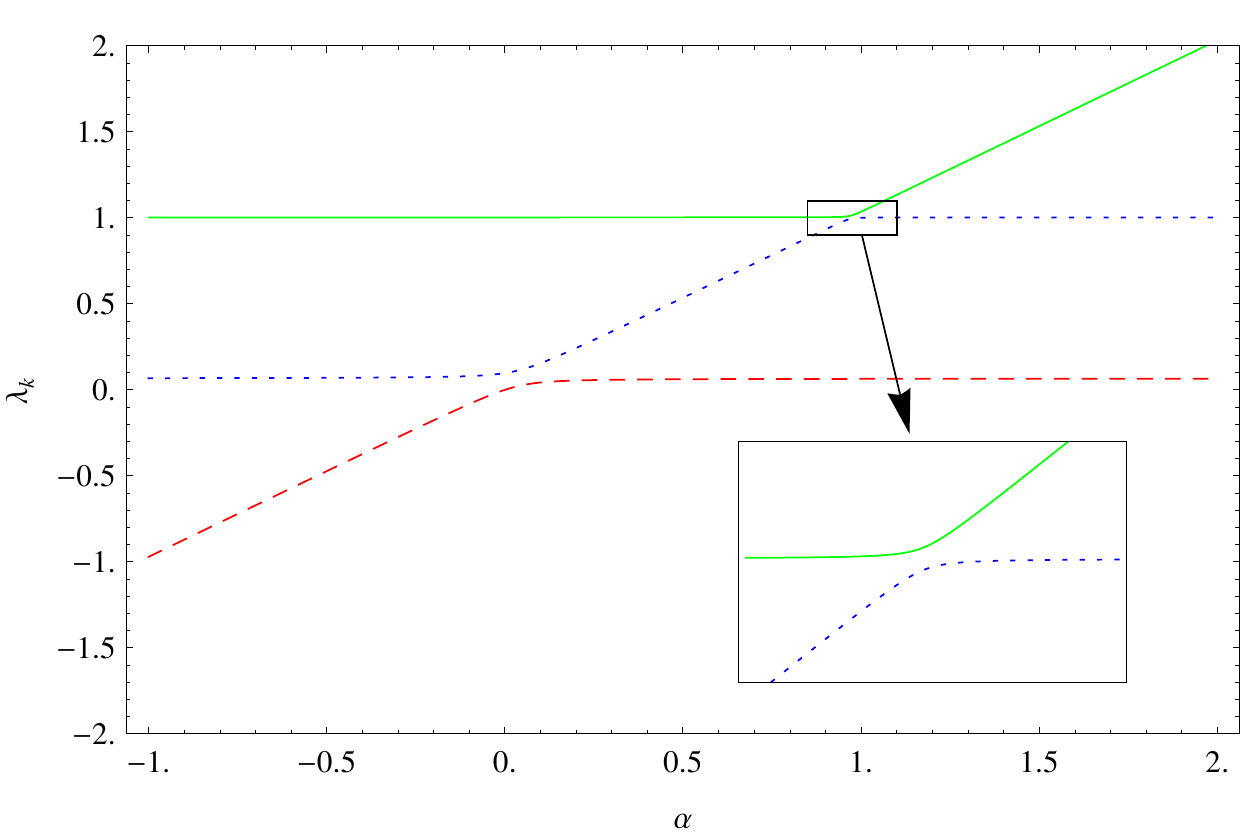}\,\includegraphics[width=7.26cm]{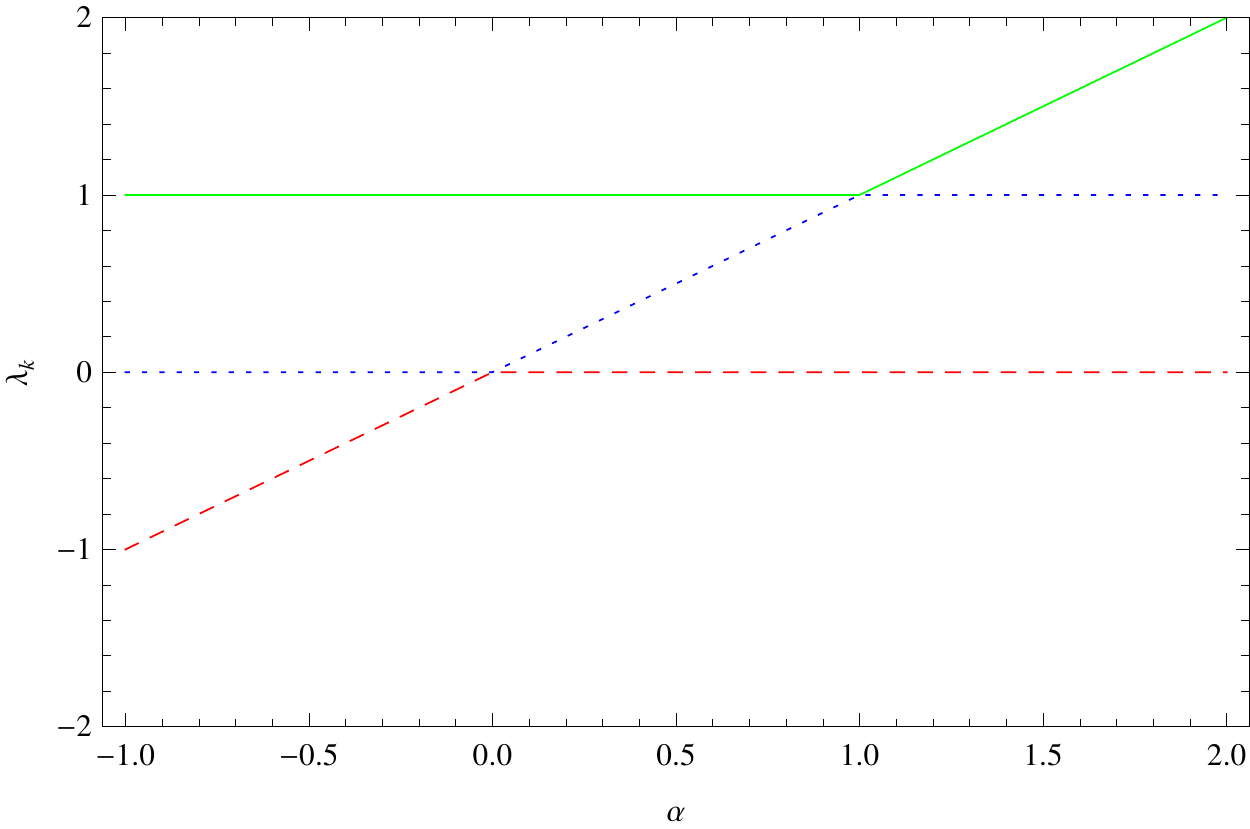}

\protect\caption{\label{fig:eigenvalues1}The three eigenvalues given by eq.(\ref{eq:1228-3})
as a function of $\alpha$. In the left plot, the matter effect parameter
is $A=0.1$, corresponding to $E=1.2$GeV. In the right plot we take
the limit $A\rightarrow0$. }
\end{figure}

First we look at the functions $\lambda_{k}(\alpha)$ which
are defined by the exact solution of eigenvalues (\ref{eq:1228-3})
and vary with $\alpha$, as shown in figure \ref{fig:eigenvalues1}.
Note that we consider $\lambda$'s as functions of $\alpha$ rather
than $A$ (or $E$), since they are expanded with respect to $\alpha$.

The left plot in figure \ref{fig:eigenvalues1} shows that the eigenvalues
can be very close to another at level crossings (corresponding to resonances),
but they never really go across another. They turn to   different
directions at   level crossings which implies the behavior near the resonance
is quite non-perturbative. 

As we suppress $A$ close to zero, the curvatures at
those turning points in figure \ref{fig:eigenvalues1} become larger
and larger. Finally the curvatures go to infinite when $A\rightarrow0$,
which makes the curves turn suddenly at some points, then some singularities
emerge. The right plot in figure \ref{fig:eigenvalues1} shows the
$A\rightarrow0$ limit. In this limit, the eigenvalues are continuous
but not differentiable functions of $\alpha$.

\begin{figure}
\centering

\includegraphics[width=7.5cm]{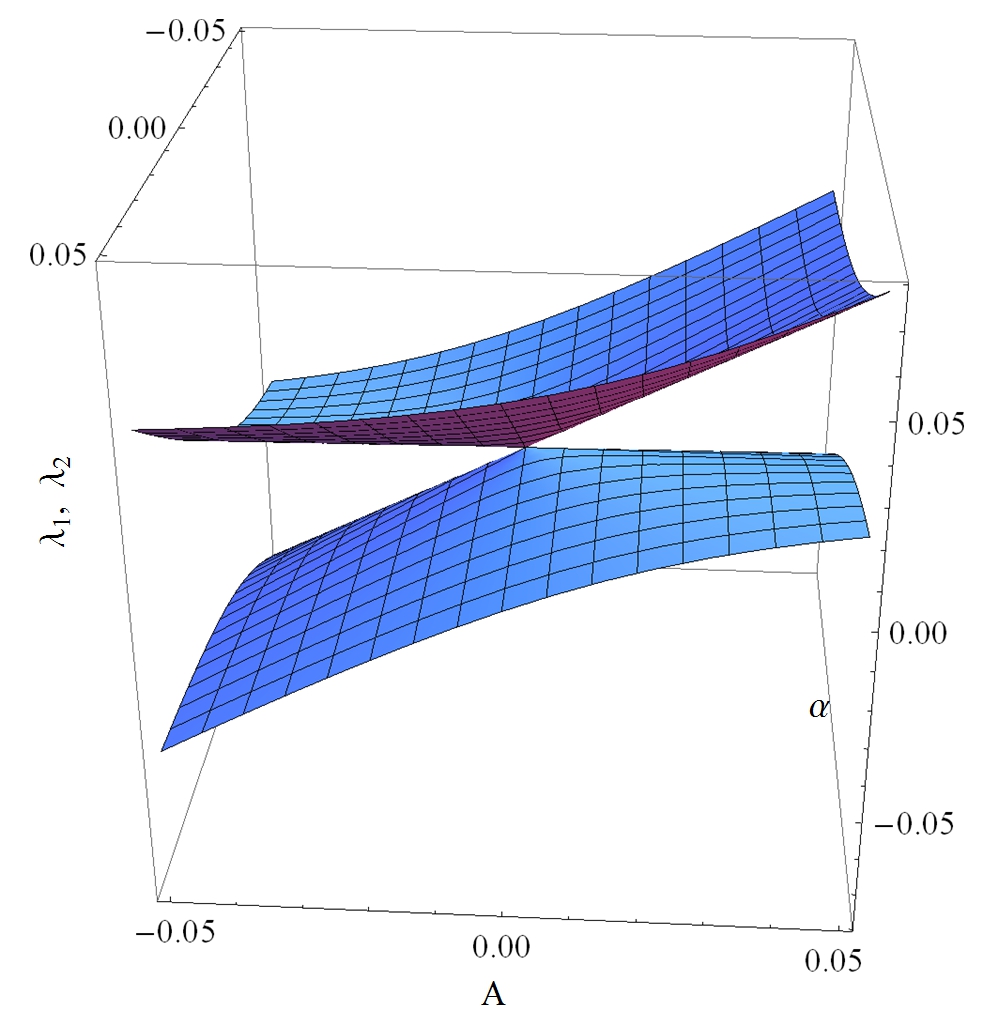}\includegraphics[width=7.5cm]{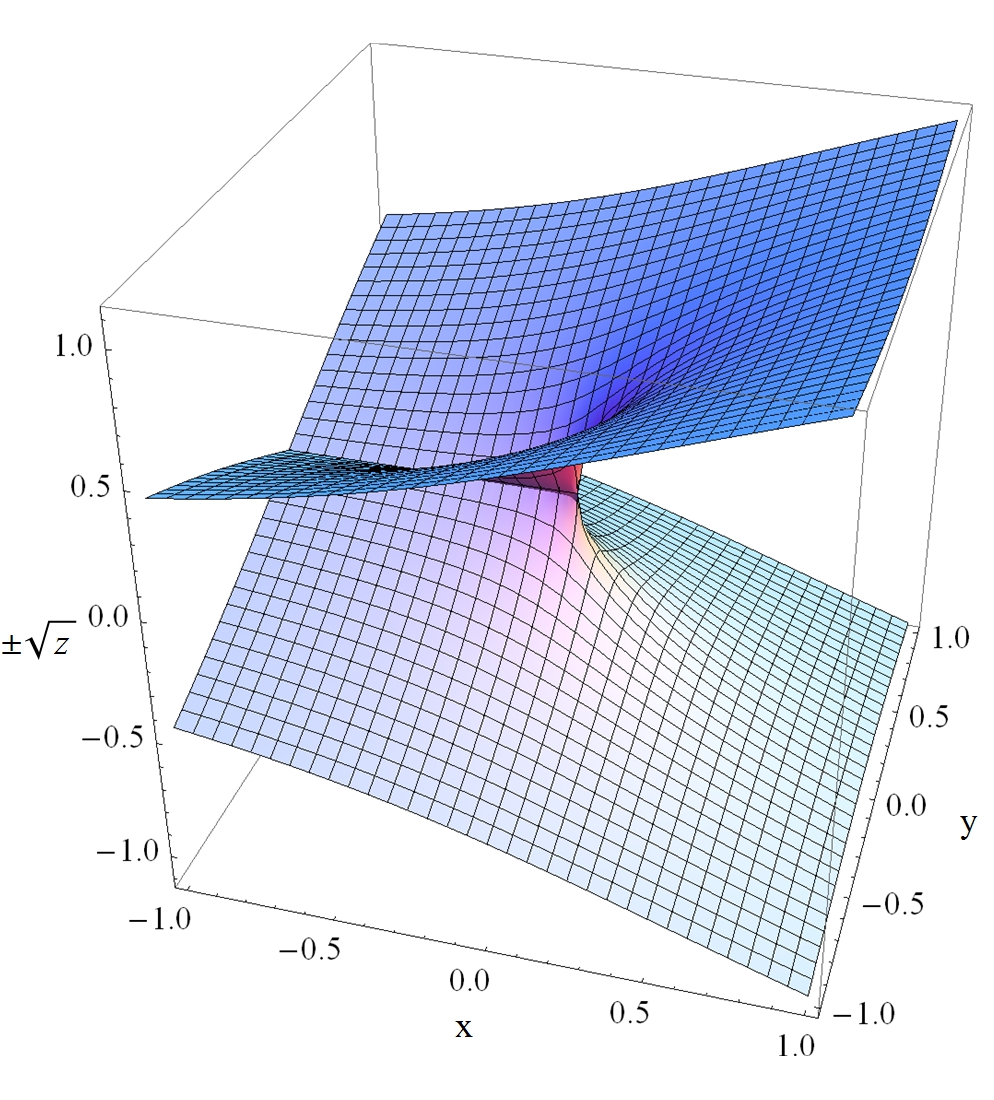}

\protect\caption{\label{fig:singular}Singularity in  the eigenvalues and its origin
from branch cut singularity. In the left plot the eigenvalues $\lambda_{1}$
and $\lambda_{2}$ from (\ref{eq:1228-3}), as functions of $\alpha$
and $A$, have a singularity at (0,0).  The singularity originates
from the branch cut in (\ref{eq:1228-6}). In the right plot we show
the branch cut singularity of $\pm\sqrt{z}$ where $z=x+iy$.}
\end{figure}

In the left plot of figure \ref{fig:singular}, we  plot
$\lambda_{1}$ and $\lambda_{2}$ as functions of $\alpha$ and $A$.
It shows that there is a singularity (here we mean non-differentiable singularity) in the eigenvalues. 
The singularity is intrinsic and can not be removed by proper
ordering of eigenvalues. So this implies  that double expansion
in $\alpha$ and $A$ does not work. 

{} The intrinsic singularity in figure \ref{fig:singular} is the kernel
of the non-perturbativity problem in the oscillation system. It actually
comes from a branch cut singularity. From eqs. (\ref{eq:1228-3})
and (\ref{eq:1228-6}) we see that $\lambda_{1,2,3}$ can be analytically
expressed in terms of $b$, $c$ and $d$ and then further in terms
of $\alpha$, $A$ and $u$ according to (\ref{eq:1228-5}). They look
like  smooth analytic functions everywhere but they are actually not. Note that
there is a square root and a cubic root in (\ref{eq:1228-6}).  Functions like $\sqrt{z}$ or $z^{1/3}$ have branch cut singularities  which make them not analytic\footnote{A complex function $g(z)$ is analytic if and only if its Taylor series
about $z_{0}$ converges to the function in some neighborhood for
every $z_{0}$ in its domain.
} on the complex plane.

In the right plot in figure \ref{fig:singular} we show the two
branches of $\pm\sqrt{z}$ where $z\equiv x+iy$ connecting
with each other at the branch cut (the line $y=0$ for $x<0$). 
At $x=0$ which is the end of the branch cut,  there is a singularity.
As one can check from (\ref{eq:1228-3}) and (\ref{eq:1228-6}),
the branch cut singularity just corresponds to the singularity in
the eigenvalues shown in the left plot. %

\section{\label{sec:Solution}Solution}

Identifying that the singularity in the eigenvalues originates from
the branch cut singularity makes a crucial step to solve the problem,
since all branch cut singularities can be easily removed if the multi-branches
collapse to one. For example, the branch cut singularity in $\sqrt{z}$
will disappear when it is squared, i.e. $(\sqrt{z})^{2}$ makes the
two branches collapse and is singularity-free. Once the singularity
is removed,   $\alpha/A$ 
will not appear any more because $1/A$ will be absorbed
by some continuous and smooth functions which we will see below. After
the singularity problem is solved, we will mathematically show  the
conventional formula  is accurate near the solar resonance.

The solution of the singularity problem can be summarized by the three key steps below: 
\begin{enumerate}
\item Use the Cayley-Hamilton theorem\cite{moler1978nineteen,Ohlsson:1999xb,Ohlsson:1999um}
to express the $S$-matrix only in terms of the eigenvalues $\lambda_{1,2,3}$. The eigenvectors are
not needed. 
\item The singularity still exists in the eigenvalues but can be partially
removed by the transformation 
\begin{equation}
\lambda_{\pm}=\frac{1}{2}(\lambda_{1}\pm\lambda_{2}),\label{eq:0107-2}
\end{equation}
where it will be shown that $\lambda_{+}$ and $\lambda_{-}^{2}$
are singularity-free, though the singularity still exists in $\lambda_{-}$
since it is a branch cut singularity. %

\item It turns out that $\lambda_{-}$ only appears in cosine function and the
following function
\begin{equation}
f(x)\equiv\frac{\sin x}{x},\label{eq:0107}
\end{equation}
where $x$ is proportional to $\lambda_{-}$. Note that $f(x)$ is smooth everywhere, even at $x=0$. Since $f(x)=1-\frac{x^{2}}{6}+\frac{x^{4}}{120}+...$
and $\cos x=1-\frac{x^{2}}{2}+\frac{x^{4}}{24}+...$ are actually
functions of $x^{2}\propto \lambda_{-}^{2}$ which is singularity-free, 
the singularity is removed. 
\end{enumerate}
In short, we will first make the $S$-matrix only depend on $(\lambda_{1},\lambda_{2},\lambda_{3})$
and then after the transformation from $(\lambda_{1},\lambda_{2},\lambda_{3})$
to $(\lambda_{-}^{2},\lambda_{+},\lambda_{3})$, the singularity in
the $S$-matrix will be explicitly removed. Next we will show the 
calculation in detail. 

The Cayley-Hamilton theorem is a theorem in linear algebra
which states that if $p(\lambda)$ is the characteristic polynomial
of a matrix $A$ [for example the left-handed side of eq.(\ref{eq:1228-4})],
then substituting the matrix $A$ for
$\lambda$ in this polynomial results in the zero matrix, i.e. $p(A)=0.$
Take the example of eq.(\ref{eq:1228-4}), this means $M^{3}+bM^{2}+cM+d=0$
or 
\begin{equation}
M^{3}=-(bM^{2}+cM+d).\label{eq:0107-3}
\end{equation}
This implies $e^{M}=I+M+M^{2}/2!+...$ can be expressed by a polynomial
of $M$ with finite terms since all $M^{n}$ with $n\geq3$ can be
converted to linear  combinations of $I,M,M^{2}$ by eq.(\ref{eq:0107-3}).
So we have
\begin{equation}
e^{-itM}=s_{0}I+s_{1}M+s_{2}M^{2},\label{eq:0101}
\end{equation}
where we put a $-it$ to be used later. The coefficient $s_{0}$,
$s_{1}$ and $s_{2}$ can be determined in various methods\cite{moler1978nineteen}
such as the Lagrange interpolation or the Newton interpolation. They
have been computed in \cite{Ohlsson:1999xb,Ohlsson:1999um}, 
\begin{eqnarray}
s_{0} & = & \frac{-1}{\Delta_{\lambda}}[e^{-it\lambda_{3}}\lambda_{1}\lambda_{2}(\lambda_{1}-\lambda_{2})+e^{-it\lambda_{1}}\lambda_{2}\lambda_{3}(\lambda_{2}-\lambda_{3})+e^{-it\lambda_{2}}\lambda_{1}\lambda_{3}(\lambda_{3}-\lambda_{1})],\label{eq:0101-1}\\
s_{1} & = & \frac{1}{\Delta_{\lambda}}[e^{-it\lambda_{3}}(\lambda_{1}^{2}-\lambda_{2}^{2})+e^{-it\lambda_{1}}(\lambda_{2}^{2}-\lambda_{3}^{2})+e^{-it\lambda_{2}}(\lambda_{3}^{2}-\lambda_{1}^{2})],\label{eq:0101-2}\\
s_{2} & = & \frac{-1}{\Delta_{\lambda}}[e^{-it\lambda_{3}}(\lambda_{1}-\lambda_{2})+e^{-it\lambda_{1}}(\lambda_{2}-\lambda_{3})+e^{-it\lambda_{2}}(\lambda_{3}-\lambda_{1})],\label{eq:0101-3}
\end{eqnarray}
where
\begin{equation}
\Delta_{\lambda}\equiv(\lambda_{1}-\lambda_{2})(\lambda_{2}-\lambda_{3})(\lambda_{3}-\lambda_{1}).\label{eq:0101-4}
\end{equation}
From eq.(\ref{eq:1228}) and (\ref{eq:1228-2}), the $S$-matrix is
\begin{equation}
S=e^{-iHL}=e^{-i\frac{m_{1}^{2}L}{2E}}e^{-2i\Delta M},\label{eq:0108}
\end{equation}
where 
\begin{equation}
M=U\left(\begin{array}{ccc}
0\\
 & \alpha\\
 &  & 1
\end{array}\right)U^{\dagger}+\left(\begin{array}{ccc}
A\\
 & 0\\
 &  & 0
\end{array}\right).\label{eq:0108-1}
\end{equation}
 So we can identify $t=2\Delta$ to use eq.(\ref{eq:0101}) directly.

Now the transition amplitude of $\nu_{\mu}\rightarrow\nu_{e}$ is
$S_{e\mu}=s_{0}I_{12}+s_{1}M_{12}+s_{2}(M^{2})_{12}$ but because
$I$ is an identity matrix, $S_{e\mu}$ can be written as two terms
\begin{equation}
S_{e\mu}=s_{1}M_{12}+s_{2}(M^{2})_{12},\label{eq:0108-4}
\end{equation}
which implies we do not need to compute $s_{0}$ for the appearance
probability.

The denominator $\Delta_{\lambda}$ in $s_{1}$ and $s_{2}$
is possible to be zero, but we will show next that $s_{1}$ and $s_{2}$
are not divergent at $\Delta_{\lambda}=0$ and smooth (differentiable)
everywhere. For example the singularity from $\lambda_{1}-\lambda_{2}=0$
can be removed by the transformation $\lambda_{\pm}=\frac{1}{2}(\lambda_{1}\pm\lambda_{2})$.
This singularity is the only one that confronts us in the energy range
of current experiments.

According to Vieta's formulas for a cubic equation, 
\begin{equation}
\lambda_{1}+\lambda_{2}+\lambda_{3}=-b,\thinspace\lambda_{1}\lambda_{2}\lambda_{3}=-d,\label{eq:0108-2}
\end{equation}
we have $\lambda_{1}+\lambda_{2}=-b-\lambda_{3}$ and $\lambda_{1}\lambda_{2}=-d\lambda_{3}^{-1}$.
Therefore
\begin{equation}
\lambda_{+}=\frac{1}{2}(-b-\lambda_{3}),\thinspace\lambda_{-}^{2}=\lambda_{+}^{2}+d\lambda_{3}^{-1},\label{eq:0108-3}
\end{equation}
where $b$, $d$ are apparently free from the singularity [as shown
in eq.(\ref{eq:1228-5})] and $\lambda_{3}$ is also singularity-free 
(shown in figure \ref{fig:eigenvalues1}, see also the proof in
the Appendix). So $\lambda_{+}$ and $\lambda_{-}^{2}$ are singularity-free.
Note that, however, $\lambda_{-}$ has a singularity originating from the branch cut singularity, which can be seen from figure \ref{fig:singular}.

After the transformation $\lambda_{\pm}=\frac{1}{2}(\lambda_{1}\pm\lambda_{2})$,
$s_{1}$ and $s_{2}$ are given by
\begin{equation}
s_{1}=\frac{-2\lambda_{+}e^{-it\lambda_{3}}+e^{-it\lambda_{+}}[2\lambda_{+}\cos(\lambda_{-}t)+it(\lambda_{+}^{2}+\lambda_{-}^{2}-\lambda_{3}^{2})f(\lambda_{-}t)]}{\lambda_{3}^{2}-2\lambda_{+}\lambda_{3}-d\lambda_{3}^{-1}},\label{eq:0108-6}
\end{equation}
\begin{equation}
s_{2}=\frac{e^{-it\lambda_{3}}+e^{-it\lambda_{+}}[-\cos(\lambda_{-}t)+it(\lambda_{3}-\lambda_{+})f(\lambda_{-}t)]}{\lambda_{3}^{2}-2\lambda_{+}\lambda_{3}-d\lambda_{3}^{-1}}.\label{eq:0108-7}
\end{equation}
We see they depend only on $\lambda_{3},$ $\lambda_{+}$ , $\lambda_{-}$
and $d$ which are all continuous and smooth functions except for
$\lambda_{-}$. But since $\lambda_{-}$ only appears in $\cos(\lambda_{-}t)$
and $f(\lambda_{-}t)$ which are actually functions of $\lambda_{-}^{2}$
( note that $\cos(x)=1-\frac{x^{2}}{2}+\frac{x^{4}}{24}+...$ and
$f(x)=1-\frac{x^{2}}{6}+\frac{x^{4}}{120}+...$), we come to the conclusion
that $s_{1}$ and $s_{2}$ are continuous and smooth functions of
$\alpha$ and $A$. 

Now that we have expressed the $S$-matrix in terms of 
$(\lambda_{3},\lambda_{+},\lambda_{-}^2)$ and thus
removed the singularity, expansion in $\alpha$ will
not suffer from any problems. The $\alpha/A$ appears in section \ref{sec:Accurarcy-of--expansion}
will not appear any more if we use eqs.(\ref{eq:0108-6}-\ref{eq:0108-7})
to compute the probability. 

Define 
\begin{equation}
p=U_{e3}U_{\mu3}^{*},\thinspace q=U_{e2}U_{\mu2}^{*},\label{eq:0108-5}
\end{equation}
we have 
\begin{equation}
S_{e\mu}=p[s_{1}+s_{2}(1+A)]+q\alpha[s_{1}+s_{2}(\alpha+A)].\label{eq:0108-8}
\end{equation}
We call the two terms in eq.(\ref{eq:0108-8}) as $p$ term and $q$
term respectively. From eqs.(\ref{eq:0108-6}-\ref{eq:0108-7}) we
have
\begin{eqnarray}
p\textrm{ term} & = & \frac{p}{\lambda_{3}^{2}-2\lambda_{+}\lambda_{3}-d\lambda_{3}^{-1}}\left[e^{-it\lambda_{3}}(\lambda_{3}-\alpha)-e^{-it\lambda_{+}}\cos(\lambda_{-}t)(\lambda_{3}-\alpha)\right.\nonumber \\
 &  & \left.+ite^{-it\lambda_{+}}f(\lambda_{-}t)(\lambda_{+}\lambda_{3}+\alpha\lambda_{+}-d-\alpha\lambda_{3})\right],\label{eq:0201}
\end{eqnarray}
\begin{eqnarray}
q\textrm{ term} & = & \frac{q\alpha}{\lambda_{3}^{2}-2\lambda_{+}\lambda_{3}-d\lambda_{3}^{-1}}\left[e^{-it\lambda_{3}}(\lambda_{3}-1)-e^{-it\lambda_{+}}\cos(\lambda_{-}t)(\lambda_{3}-1)\right.\nonumber \\
 &  & \left.-ite^{-it\lambda_{+}}f(\lambda_{-}t)(\lambda_{3}^{2}-2\lambda_{+}\lambda_{3}-d\lambda_{3}^{-1}+(\lambda_{3}-1)(\lambda_{3}-\lambda_{+}))\right].\label{eq:0201-1}
\end{eqnarray}

So far we have not taken any approximation. 
Then we will use the approximation
\begin{equation}
\lambda_{3}=1+\mathcal{O}(s_{13}^{2}A),\label{eq:0123-4}
\end{equation}
which is derived in the Appendix. With this approximation, from
eq.(\ref{eq:0108-3}) we have 

\begin{equation}
2\lambda_{+}=A+\alpha+\mathcal{O}(s_{13}^{2}A),\thinspace\lambda_{-}^{2}=\lambda_{+}^{2}-\frac{A\alpha c_{12}^{2}c_{13}^{2}}{1+\mathcal{O}(s_{13}^{2}A)}.\label{eq:0205-2}
\end{equation}
Since the $p$ term and $q$ term have been expressed in terms of
singularity-free quantities $\lambda_{3}$, $\lambda_{+}$ and $\lambda_{-}^{2}$,
we can use (\ref{eq:0123-4},\ref{eq:0205-2}) to compute them. The
calculation is straightforward (see the Appendix) and the result is
\begin{equation}
p\textrm{ term}=p\frac{e^{-2i\Delta}-e^{-2iA\Delta}}{1-A}+\mathcal{O}(\Delta s_{13}^{3}A)+\mathcal{O}(\Delta^{2}s_{13}\alpha A),\label{eq:0108-11}
\end{equation}
\begin{equation}
q\textrm{ term}=-2iq\alpha e^{-i(A+\alpha)\Delta}\frac{\sin(\bar{A}\Delta)}{\bar{A}}+\mathcal{O}(\Delta\alpha s_{13}^{2}A),\label{eq:0108-12}
\end{equation}
where
\begin{equation}
\bar{A}\equiv\sqrt{(A+\alpha)^{2}-4A\alpha c_{12}^{2}c_{13}^{2}}.\label{eq:0112}
\end{equation}
The  oscillation probability is 
\begin{eqnarray}
 &  & |p\textrm{ term}+q\textrm{ term}|^{2}\nonumber \\
 & = & |p\frac{e^{-2i\Delta}-e^{-2iA\Delta}}{1-A}-2iq\alpha e^{-i(A+\alpha)\Delta}\frac{\sin(\bar{A}\Delta)}{\bar{A}}+\mathcal{O}(\Delta s_{13}^{3}A)+\mathcal{O}(\Delta^{2}s_{13}\alpha A)|^{2}\nonumber \\
 & = & P^{(A)}+\mathcal{O}(s_{13}^{4}A\Delta)+\mathcal{O}(s_{13}^{2}\alpha A\Delta^{2}),\label{eq:0207-1}
\end{eqnarray}
where $P^{(A)}$ is defined as 
\begin{equation}
P^{(A)}=|p\frac{e^{-2i\Delta}-e^{-2iA\Delta}}{1-A}-2iq\alpha e^{-i(A+\alpha)\Delta}\frac{\sin(\bar{A}\Delta)}{\bar{A}}|^{2}.\label{eq:0112-1}
\end{equation}
To derive (\ref{eq:001}), we need to expand
the modulus squared of (\ref{eq:0112-1}), 
\begin{equation}
P^{(A)}=|p\textrm{ term}|^{2}+2\textrm{Re}[\overline{p\textrm{ term}}\times q\textrm{ term}]+|q\textrm{ term}|^{2},\label{eq:0829-2}
\end{equation}
where $|p\textrm{ term}|^{2}$, $|q\textrm{ term}|^{2}$ and the cross
term are computed in the appendix and the result is  
\begin{equation}
|p\textrm{ term}|^{2}=4s_{13}^{2}c_{13}^{2}s_{23}^{2}\frac{\sin^{2}(1-A)\Delta}{(1-A)^{2}},\label{eq:0829}
\end{equation}
\begin{equation}
|q\textrm{ term}|^{2}=4\alpha^{2}s_{12}^{2}c_{12}^{2}c_{23}^{2}\frac{\sin^{2}(A\Delta)}{A^{2}}+\mathcal{O}(\alpha^{4}\Delta^{4})+\mathcal{O}(A\alpha^{3}\Delta^{4})+\mathcal{O}(\alpha^{2}s_{13}\Delta^{2})),\label{eq:0829-1}
\end{equation}
\begin{equation}
2\textrm{Re}[\overline{p\textrm{ term}}\times q\textrm{ term}]=8\alpha\frac{J_{CP}}{s_{\delta}}\cos(\Delta+\delta)\frac{\sin A\Delta}{A}\frac{\sin(1-A)\Delta}{1-A}+\mathcal{O}(\alpha s_{13}^{2}\Delta)+\mathcal{O}(s_{13}\alpha^{2}\Delta).
\label{eq:0829-4}
\end{equation}


Now the conventional formula (\ref{eq:001}) %
can be analytically justified near and below the solar resonance.
Here we denote it as $P^{(B)}$ . 
We can see that the first and last terms of $P^{B}$ just correspond
to eqs.(\ref{eq:0829},\ref{eq:0829-1}) respectively and the middle
term in $P^{(B)}$ corresponds to the cross term (\ref{eq:0829-4}).
Combine the analytic errors, we have
\begin{equation}
P^{(B)}-P^{(A)}=\mathcal{O}(s_{13}^{2}\alpha\Delta)+\mathcal{O}(s_{13}\alpha^{2}\Delta^{2})+\mathcal{O}(\alpha^{3}A\Delta^{4})+\mathcal{O}(\alpha^{4}\Delta^{4}),\label{eq:0207}
\end{equation}
which implies that the conventional formula is accurate up to the $\mathcal{O}$-terms
above and the $\mathcal{O}$-terms in (\ref{eq:0207-1}). We see there
is no $\alpha/A$ in all these $\mathcal{O}$'s, 
so we draw the conclusion that the bound (\ref{eq:1225-3}) which
originally requires $\alpha/A\ll1$ can be safely removed. The conventional
formula is still accurate without this bound, as long as these $\mathcal{O}$-terms
are small.

According to  eq.(\ref{eq:0207-1}), the error of $P^{(A)}$ is 
$\delta P^{(A)}=\mathcal{O}(s_{13}^{4}A\Delta)+\mathcal{O}(s_{13}^{2}\alpha A\Delta^{2})$.
Taking T2K as an example, for
$E=0.25$GeV which is below the conventional domain of validity,
we have 
 $A\simeq0.02$ and $\Delta\simeq3.7$. This gives 
$s_{13}^{2}\alpha A\Delta^{2}\simeq2\times10^{-4}$ 
and $s_{13}^{4}A\Delta\simeq4\times10^{-5}$.
The error is very small and the dominant correction 
would be $\mathcal{O}(s_{13}^{2}\alpha A\Delta^{2})$ if we want to 
improve the accuracy. Actually, if we only concern ourselves with 
the $\Delta\gtrsim1$
region, then $\mathcal{O}(s_{13}^{2}\alpha A\Delta^{2})$ is larger
than $\mathcal{O}(s_{13}^{4}A\Delta)$ since $s_{13}^{2}\alpha\simeq s_{13}^{4}$.
Therefore we expect that typically $\mathcal{O}(s_{13}^{2}\alpha A\Delta^{2})$ 
is the principal source of the 
error of $P^{(A)}$. For the same set of parameter
values, the four terms in eq.(\ref{eq:0207}) have the values, $s_{13}^{2}\alpha\Delta\simeq3\times10^{-3}$,
$s_{13}\alpha^{2}\Delta^{2}\simeq2\times10^{-3}$ and $\alpha^{3}A\Delta^{4}\simeq\alpha^{4}\Delta^{4}\simeq1\times10^{-4}$.
So the dominant  errors of $P^{(B)}$ are $\mathcal{O}(s_{13}^{2}\alpha\Delta)$
and $\mathcal{O}(s_{13}\alpha^{2}\Delta^{2})$. Note that they do
not depend on  $A$, which implies that the main source of inaccuracy of $P^{(B)}$
is not due to 
inadequately accounting for 
the matter effect contribution, 
but rather than due to an insufficient expansion of the small phase $\alpha\Delta$.
Though the phase $\alpha\Delta$ is small,  terms quadratic in $\alpha\Delta$ would not
be  negligible if we want to improve its accuracy. In conclusion,
the dominant errors of $P^{(A)}$ and $P^{(B)}$ are given by $\mathcal{O}(s_{13}^{2}\alpha A\Delta^{2})$
and $\mathcal{O}(s_{13}^{2}\alpha\Delta)+\mathcal{O}(s_{13}\alpha^{2}\Delta^{2})$, 
respectively.

\section{\label{sec:discuss} Numerical verification}

As our study of the problem is originally motivated by the fact that
T2K covers the solar resonance, we would like to numerically 
verify our analysis in that case first.
The matter
density in T2K is $\rho=2.6\textrm{g}/\textrm{cm}^{3}$\cite{T2K2013PRD}
so we take the electron density to be $N_{e}=1.3N_{A}/\textrm{cm}^{3}$
under the assumption that for matter $Z/A=1/2$  in average.

\begin{figure}
\centering

\includegraphics[width=7.1cm]{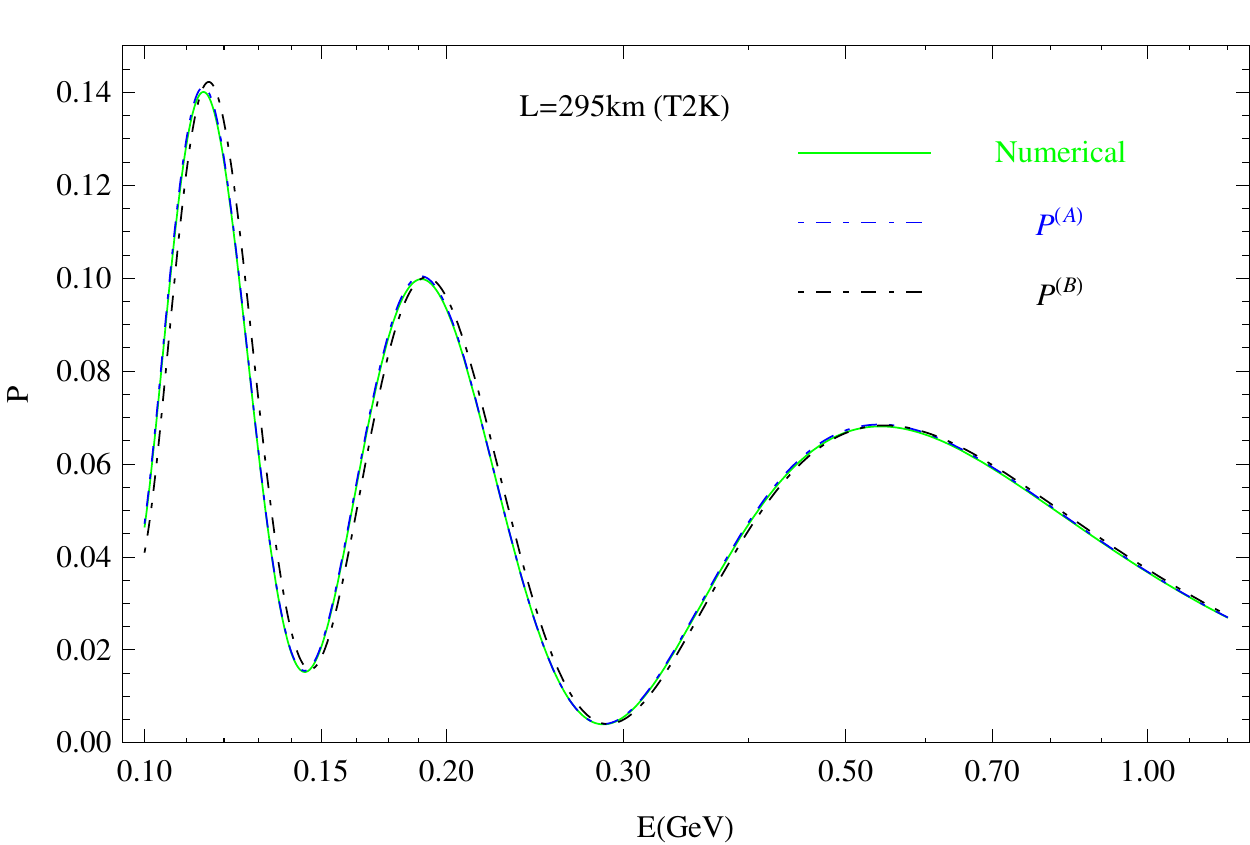}\hspace{2mm}\includegraphics[width=7.1cm]{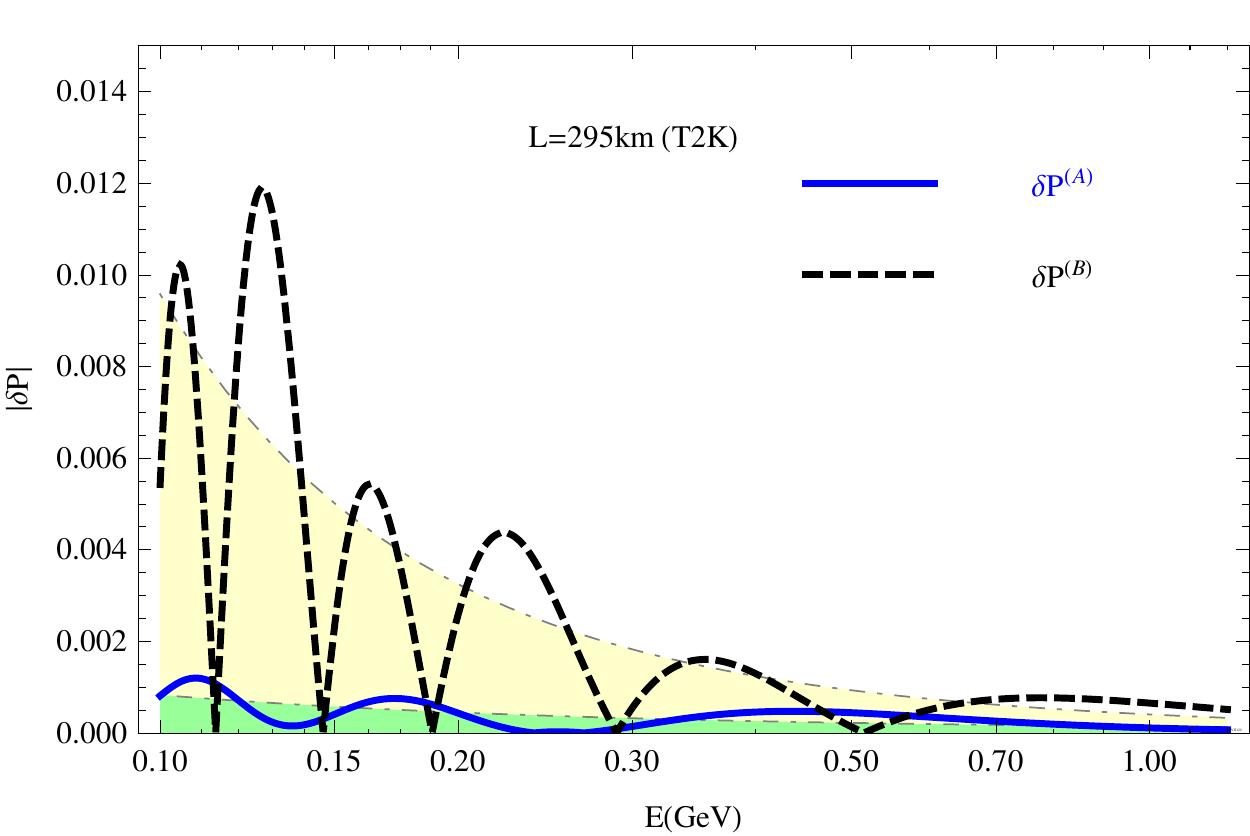}

\protect\caption{\label{fig:proba-all}A comparison of  the approximate oscillation formulae
 with the numerical solution to (\ref{eq:1228}) in the T2K
case. The left plot shows that all these approximate formulae are
very accurate and the errors are negligible for practical use. The
right plot shows that the residuals defined as $|\delta P|=|P-P_{\textrm{numerical}}|$
are consistent with our analytic estimation of the error,  which are represented
by green and yellow shades for $\delta P^{(A)}$ and $\delta P^{(B)}$
respectively.}
\end{figure}

Figure \ref{fig:proba-all} shows that both $P^{(A)}$ and $P^{(B)}$
are accurate enough for practical use while the new formula $P^{(A)}$
has
better accuracy than  the conventional formula $P^{(B)}$.
We also plot the analytic errors according to (\ref{eq:0207-1}) and (\ref{eq:0207})
in the
right panel of figure \ref{fig:proba-all}, using light green and yellow
shades. The actual residuals defined as $|\delta P|=|P-P_{\textrm{numerical}}|$
where $P_{\textrm{numerical}}$ is the  numerical solution are well compatible
with analytic estimation, which implies the errors are correctly estimated. 
Therefore figure \ref{fig:proba-all} verifies 
 both $P^{(A,B)}$ and $\delta P^{(A,B)}$ in the T2K case.

Besides T2K, we also show the accuracies of these formulae in other
accelerator neutrino experiments. The information of the baselines
and neutrino energies are listed in table \ref{tab:exp} and for simplicity
we take the same matter density as T2K for all the other experiments,
since the neutrino beams in these experiments only go though the earth
crust. We see again that in current or future accelerator neutrino
experiments, the formulae are accurate enough for practical use and
the errors are well described by our analytic estimation.

\begin{table}[h]
\centering

\begin{tabular}{|ccccc|}
\hline 
Experiments & $L$/km & $E$/GeV  & $A(E/\textrm{GeV})$ & Refs\tabularnewline
\hline 
MOMENT & 150 & $\sim0.3$ & $0.024$ & \cite{MOMENT}\tabularnewline[0.1cm]
\hline 
T2K & 295 & $0.6\thinspace(0.1\rightarrow1.2)$ & $0.048\thinspace(0.008\rightarrow0.096)$ & \cite{T2K2013PRL,T2K2014PRL}\tabularnewline[0.1cm]
\hline 
MINOS  & 735 & $3\thinspace(1\rightarrow6)$ & $0.24\thinspace(0.08\rightarrow0.48)$ & \cite{MINOS} \tabularnewline
\hline 
NOvA & 810 & $\sim2$ & $0.16$  & \cite{NOVA,NOVA2}\tabularnewline
\hline 
LBNE & 1300 & $\sim2.5$ & $0.20$ & \cite{LBNE}\tabularnewline
\hline 
\end{tabular}\protect\caption{\label{tab:exp}Baseline lengths and neutrino energies of current
and future accelerator neutrino experiments. For T2K and MINOS there
are both peaks and ranges (in parentheses) of neutrino energies according
to the references while for the other experiments we only show the
general energies. The electron density is $N_{e}=1.3N_{A}/\textrm{cm}^{3}$
in our calculation, so we also show the values of $A$ corresponding
to the energies. }
\end{table}

\begin{figure}
\centering

\includegraphics[width=7.1cm]{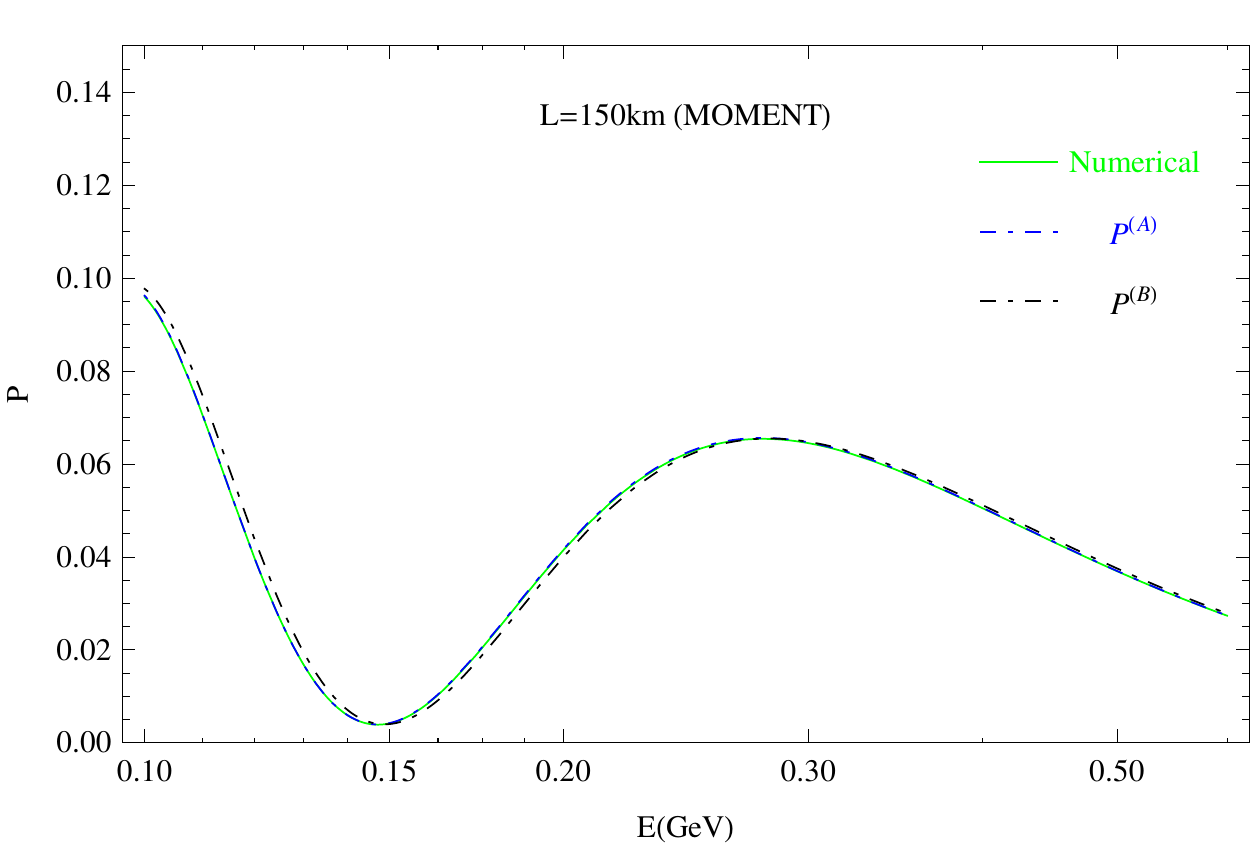}\hspace{2mm}\includegraphics[width=7.1cm]{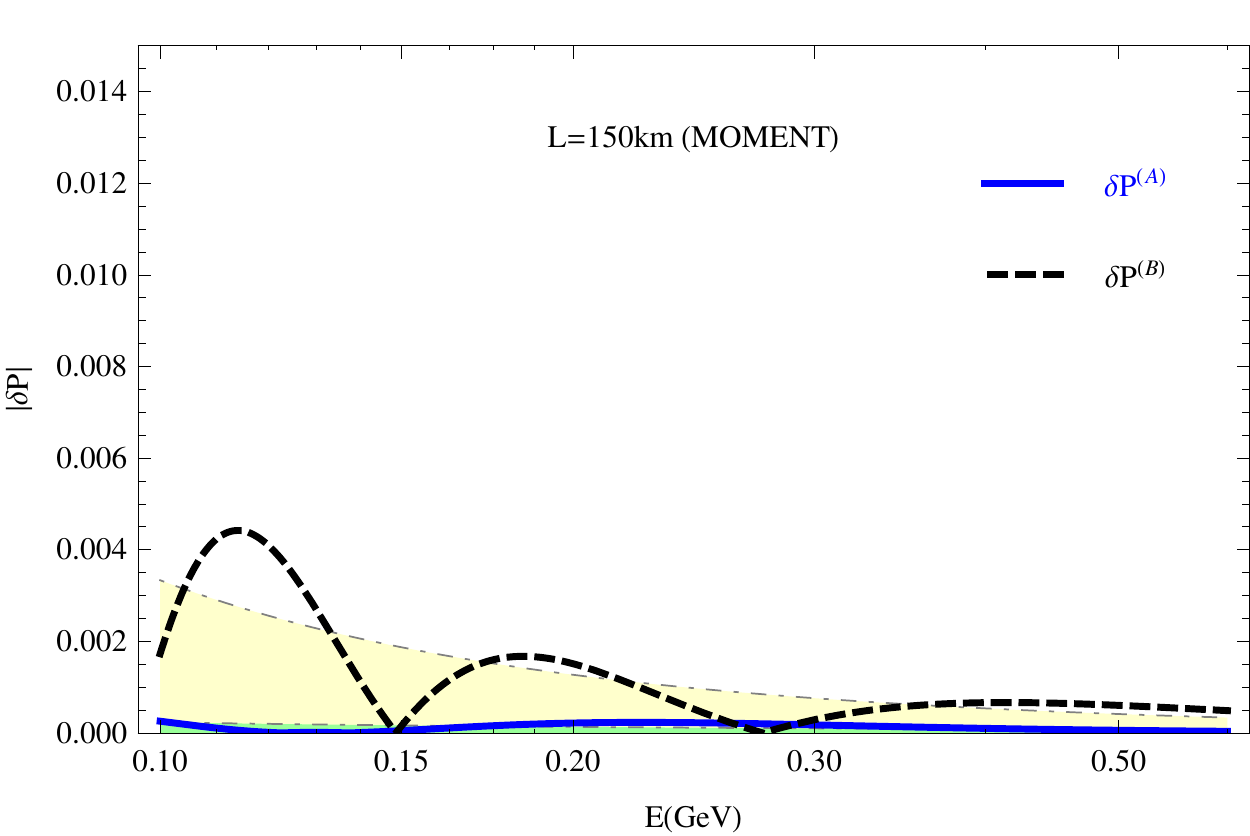}

\includegraphics[width=7.1cm]{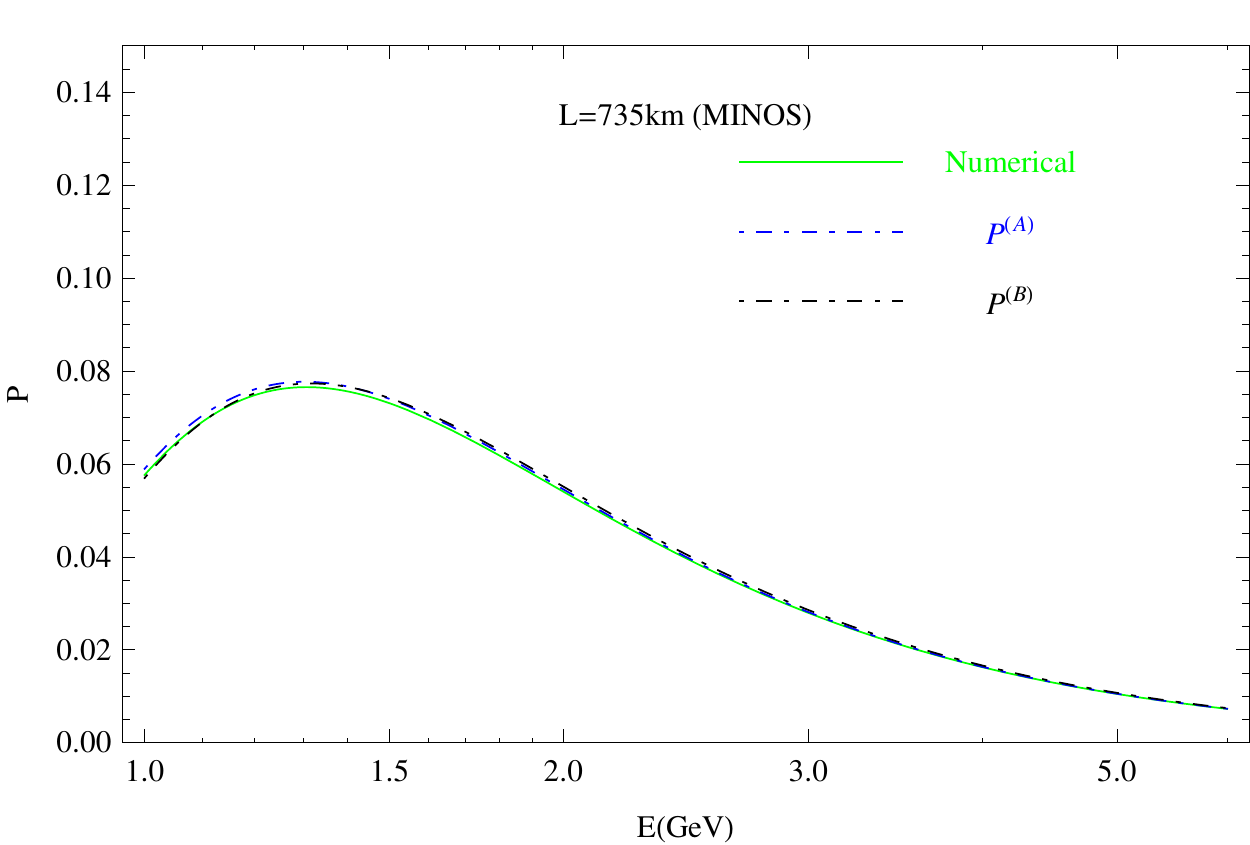}\hspace{2mm}\includegraphics[width=7.1cm]{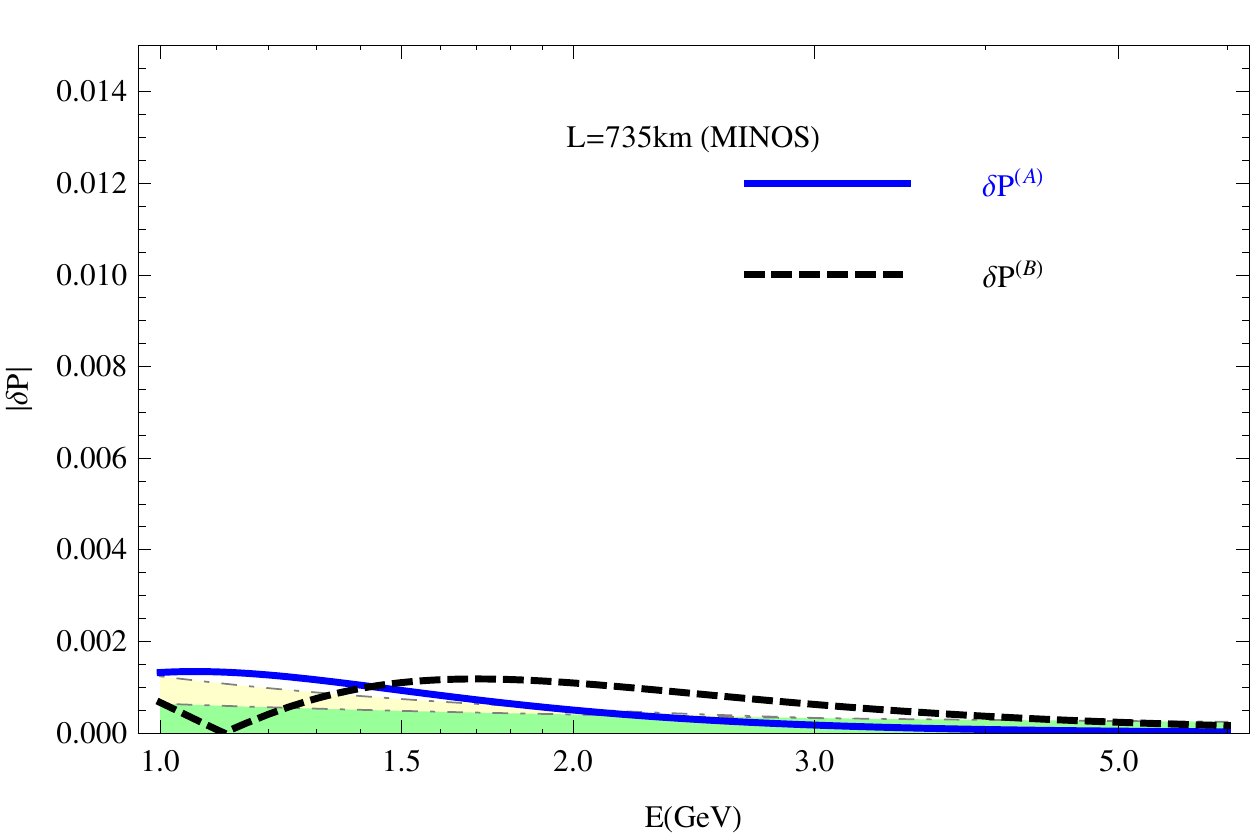}

\includegraphics[width=7.1cm]{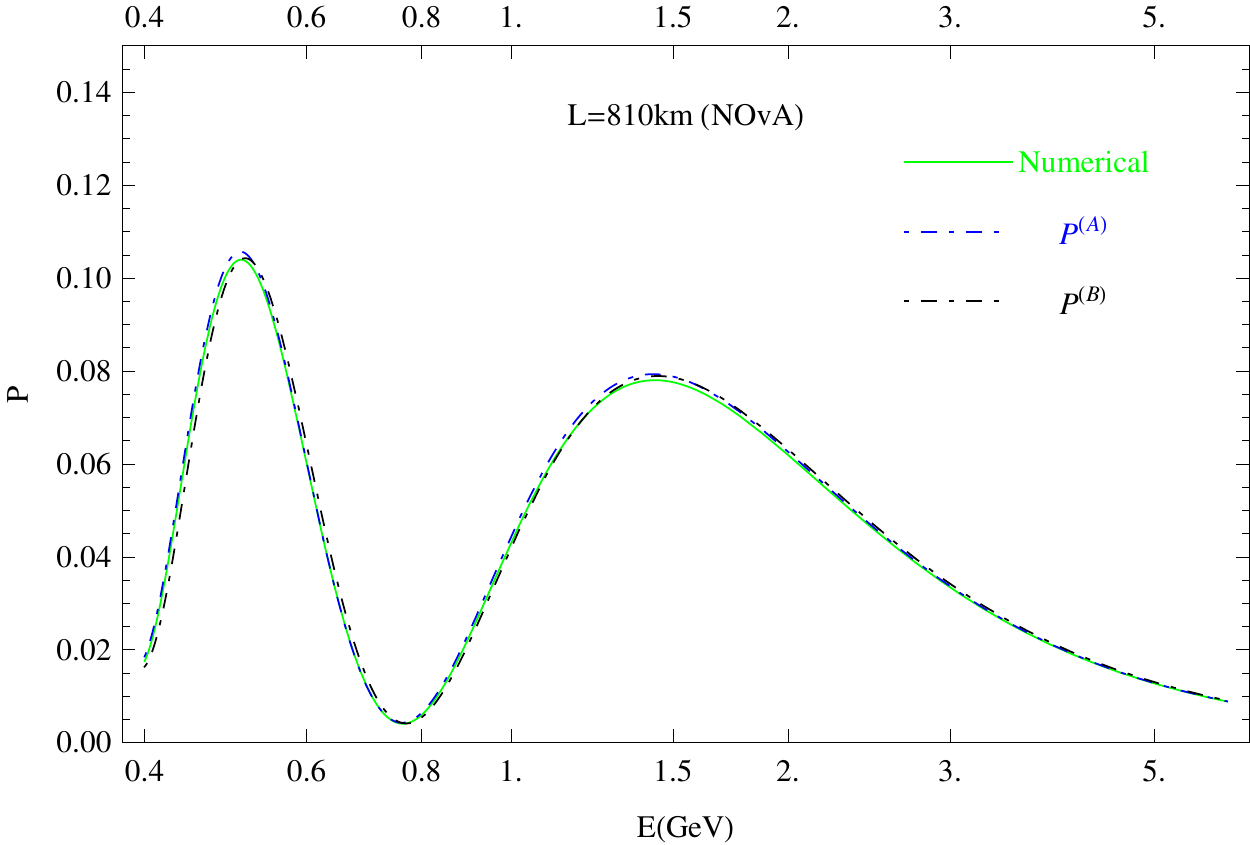}\hspace{2mm}\includegraphics[width=7.1cm]{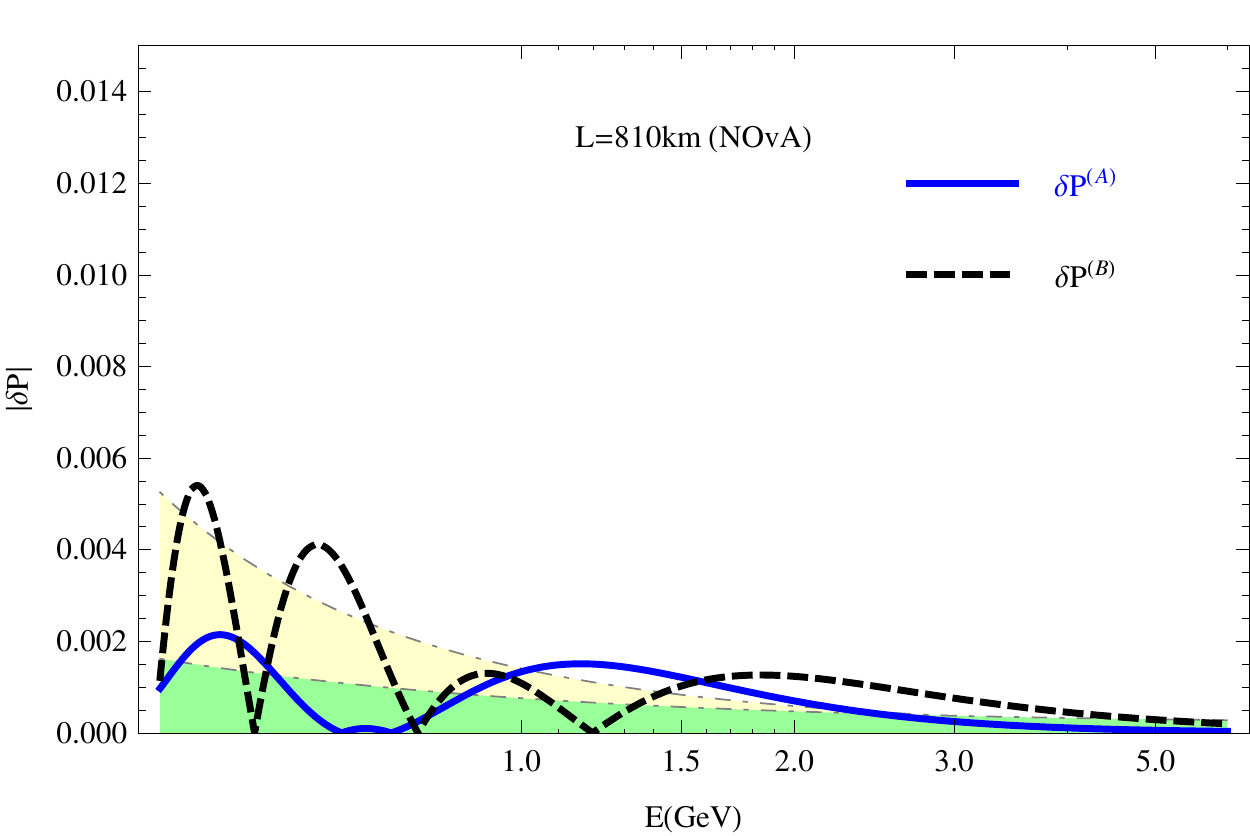}

\includegraphics[width=7.1cm]{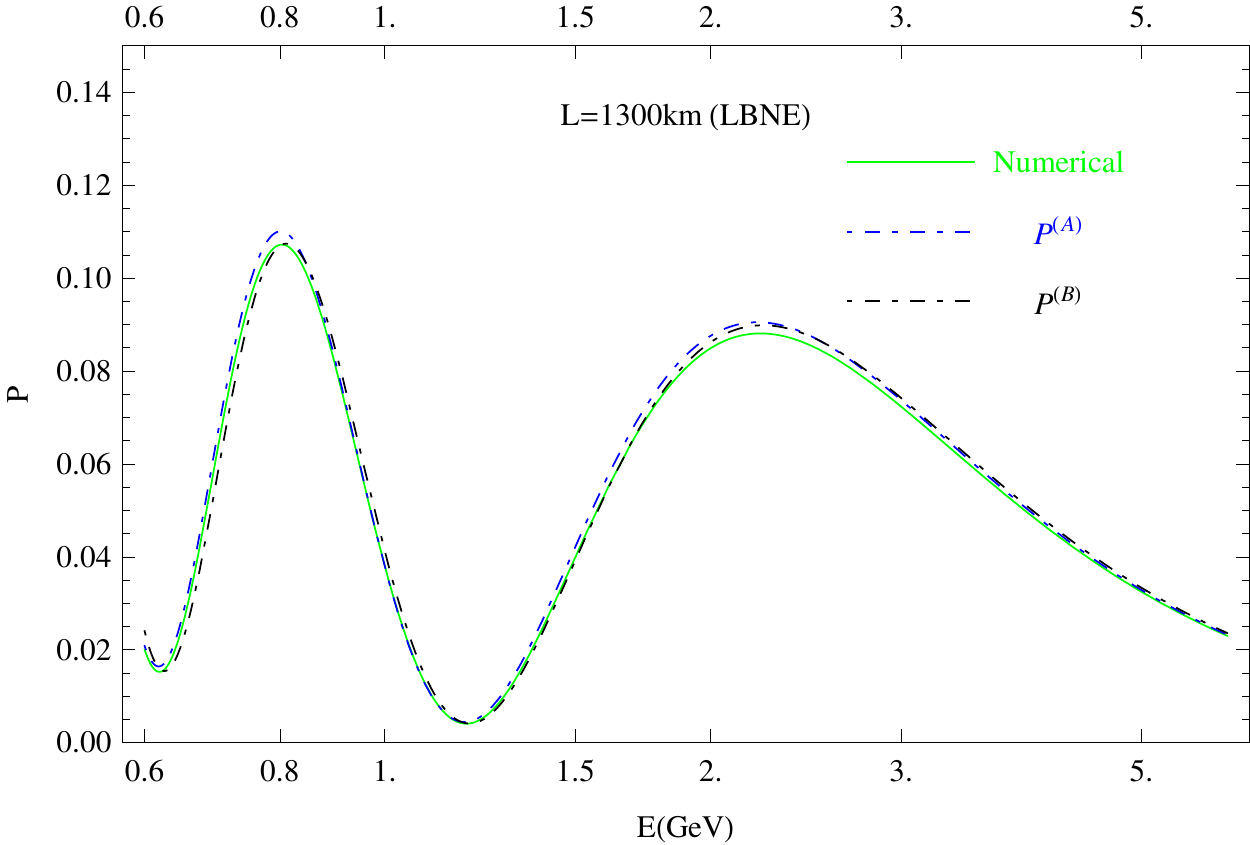}\hspace{2mm}\includegraphics[width=7.1cm]{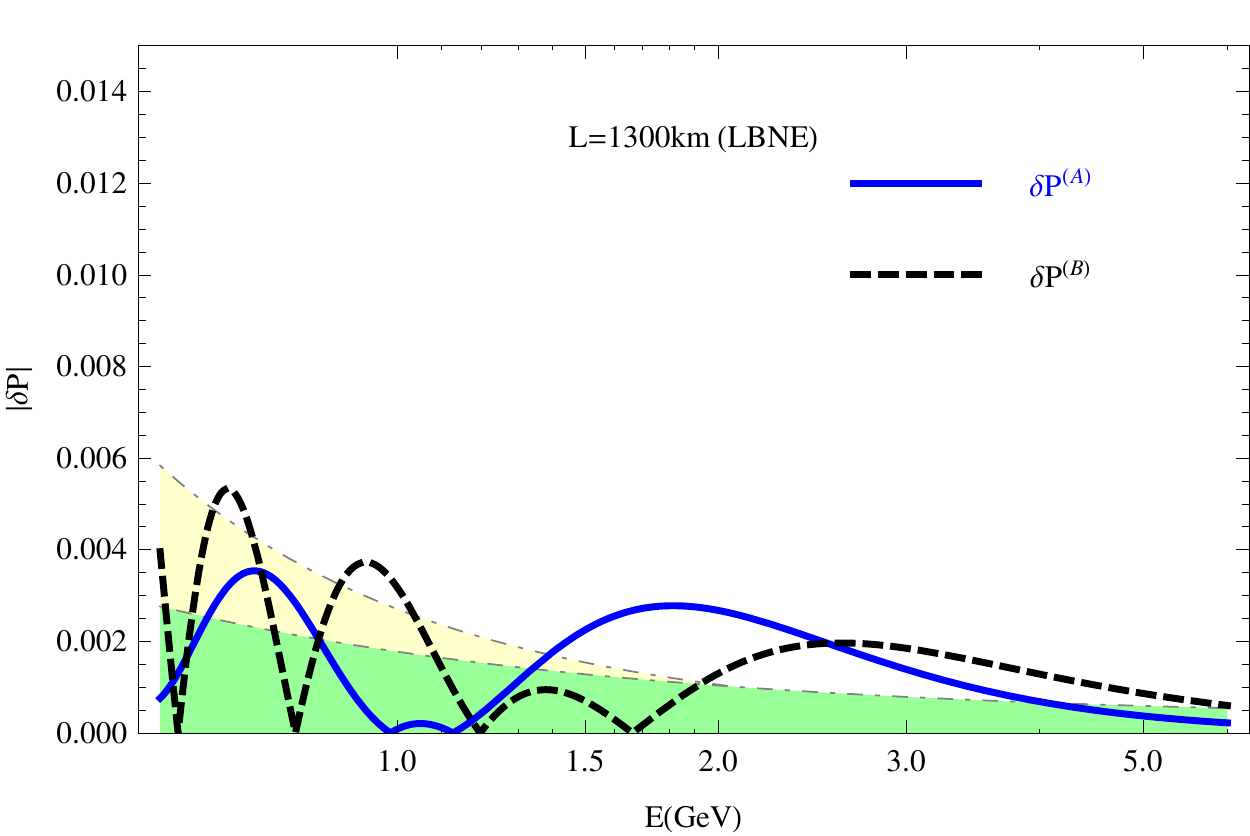}

\protect\caption{\label{fig:proba-all-1}Similar plots to figure \ref{fig:proba-all}
but for  some other accelerator neutrino experiments. For more details,
see the T2K case in figure \ref{fig:proba-all}. }
\end{figure}

Although computers are becoming more powerful, it is still 
desirable to have efficient methods of computation. 
For example, the $\chi^{2}$-fit in a high dimensional
parameter space (including both oscillation parameters and experiment
parameters) is always extremely time-consuming. When nuisance parameters
are being marginalized, the likelihood function has to be invoked
an enormous amount of times to complete
a sub-process of minimization (only for the frequentist treatment,
the Bayesian approach usually needs much more computations). The package GLoBES \cite{Huber:2004ka,Huber:2007ji}
which was designed for simulation of neutrino oscillation experiments
has optimized the diagonalization procedure \cite{Kopp:2006wp} combining
the QL decomposition with additional developed algorithms. This
is not necessary for oscillations in constant density matter, where a simple
analytical formula performs better (GLoBES allows users to replace the
probability engine with a user-defined function). In this case, we recommend 
the use of $P^{(A)}$ instead.
A simple test on Mathematica 8.0 
 with Intel Core i7 CPU  %
 shows that $10^{5}$  evaluations of $P^{(A)}$ and $P^{(B)}$
cost\footnote{For compiled languages which are used in practical simulation such
as the C-based GLoBES package, the speed will be about several hundred
times faster. But the ratio of the speeds of computing $P^{(A)}$ and $p^{(B)}$ varies little for different machines or different languages.%
} $4.7$ and $5.6$ seconds, which implies $P^{(A)}$ can be computed
at a speed not slower than $P^{(B)}$. 

Finally there is one issue related to the solar resonance to be discussed.
Strictly speaking, the matter effect can be safely regarded as a small
perturbative effect only if $2\sqrt{2}G_{F}N_{e}E$ is much less than
$\Delta m_{21}^{2}$ ($A\ll\alpha$), i.e. only the region between
the vacuum limit and the solar resonance can be regarded as the truly
small-$A$ region, where no physics can be changed greatly by the
matter effect. Typically LBL accelerator neutrino experiments are
in the region between the solar resonance and the atmospheric resonance
which we can refer to as medium-$A$ region. Note that originally
$A$ can not be treated perturbatively in the medium-$A$ region\cite{Freund:2001pn}.
From the small-$A$ region to the medium-$A$ region, the solar mixing
will experience a resonance. It is interesting that, according
to the formulae we derived , the contribution from the matter effect
passes through the resonance gradually without showing any resonances,
despite the solar mixing being affected drastically in that region 
(note that the solar mixing has sizable contributions
to these experiments). In other words, the region with a perturbative
matter correction can be extended from the small-$A$ region to the
medium-$A$ region for current LBL accelerator neutrino experiments.

\section{\label{sec:Conclusion}Conclusion}

The conventional formula obtained by an expansion in the mass hierarchy
parameter $\alpha=\Delta m_{21}^{2}/\Delta m_{31}^{2}\approx0.03$
turns out to be very accurate near the solar resonance, as shown in
figure \ref{fig:comp1} though the
effective masses and effective mixing angles computed in the $\alpha$-expansion
are inaccurate or even invalid at this region, as shown in figure
\ref{fig:mixing} and figure \ref{fig:mass}. So it is interesting
that the intermediate inaccuracies  cancel each other out
in the final result. 

We have shown that the inaccuracies are because the expansion is too
close to the branch cut singularity in the eigenvalues. This 
singularity
is inherent in the eigenvalues so it cannot be removed by interchanging
eigenvalues. But certain combinations of them such as their sum of the
eigenvalues $\lambda_{1}+\lambda_{2}$ do not have the singularity,
and the oscillation probability only depends on these singularity-free
combinations. By computing the probability in this way, we have analytically
proven that the conventional formula is still accurate near the solar
resonance.

A new oscillation formula $P^{(A)}$ in (\ref{eq:0112-1}) 
which might be practically useful 
is derived when we try to prove the 
accuracy of the conventional one. 
Both the conventional and the new formulae 
are very accurate in various accelerator
neutrino experiments for baseline lengths varying from $150$km
(MOMENT) to $1300$km (LBNE), as shown in figures \ref{fig:proba-all}
and \ref{fig:proba-all-1}. We have also  estimated the analytic errors
 for these formulae.

{} 

\appendix

\section{Some details of analytic calculations}

\subsection{Simplify the $p$ term and $q$ term}

Here we show how to simplify the $p$ term and $q$ term step by step.
All the approximations in the calculation should be analytically treated
so we use $\mathcal{O}()$ instead of $\approx$. 

The first result we will derive is eq.(\ref{eq:0123-4}). Note that
the cubic equation (\ref{eq:1228-4}) has the following identity 
\begin{equation}
b+c+d=As_{13}^{2}(\alpha-1)-1,\label{eq:0123-1}
\end{equation}
which provides a fast way to compute $\lambda_{3}$ as follow. We
assume $\lambda_{3}=1+x$ with $x\ll1$ and replace the $\lambda$
in the cubic equation (\ref{eq:1228-4}) with $1+x$. Then the leading
order vanishes and the next-to-leading order(NLO) gives 
\begin{equation}
3x+x(2b+c)+\mathcal{O}(x^{2})=As_{13}^{2}(1-\alpha),\label{eq:0123-2}
\end{equation}
which implies $x=\mathcal{O}(s_{13}^{2}A)$ while the explicit form
of $x$ is not important here. 

Therefore from eq.(\ref{eq:0108-3}) we have 

\begin{equation}
2\lambda_{+}=A+\alpha+x,\thinspace\lambda_{-}^{2}=\lambda_{+}^{2}-\lambda_{1}\lambda_{2},\thinspace\lambda_{1}\lambda_{2}=\frac{A\alpha u_{1}^{2}}{1+x},\label{eq:0205}
\end{equation}
so the $q$ term can be greatly simplified,
\begin{eqnarray}
q\textrm{ term} & = & \frac{q\alpha[\mathcal{O}(\Delta x)+ite^{-it\lambda_{+}}f(\lambda_{-}t)(\lambda_{3}^{2}-2\lambda_{+}\lambda_{3}-d\lambda_{3}^{-1}+x(\lambda_{3}-\lambda_{+}))]}{\lambda_{3}^{2}-2\lambda_{+}\lambda_{3}-d\lambda_{3}^{-1}}\nonumber \\
 & = & -itq\alpha e^{-it\lambda_{+}}f(\lambda_{-}t)+\mathcal{O}(\alpha\Delta x)\nonumber \\
 & = & -2iq\alpha e^{-i(A+\alpha+x)\Delta}\sin(\bar{A}\Delta)/\bar{A}+\mathcal{O}(\alpha\Delta s_{13}^{2}A),\label{eq:0206-13}
\end{eqnarray}
where $\bar{A}\equiv\sqrt{(A+\alpha)^{2}-4A\alpha c_{12}^{2}c_{13}^{2}}$
or $\bar{A}^{2}=4\lambda_{-}^{2}$. 
\begin{eqnarray}
p\textrm{ term} & = & \frac{p}{(1-A)(1-\alpha)+\mathcal{O}(x)+\mathcal{O}(\alpha A)}\left[e^{-it}(1-\alpha)-e^{-it\lambda_{+}}\cos(\lambda_{-}t)(1-\alpha)+\mathcal{O}(x)\right.\nonumber \\
 &  & \left.+ite^{-it\lambda_{+}}f(\lambda_{-}t)\left((1-\alpha)(A-\alpha)/2+\mathcal{O}(\alpha A)+\mathcal{O}(\alpha x)+\mathcal{O}(\lambda_{+}x)\right)\right]\nonumber \\
 & = & \frac{p\left[e^{-it}(1-\alpha)-e^{-it\lambda_{+}}\cos(\lambda_{-}t)(1-\alpha)+ite^{-it\lambda_{+}}f(\lambda_{-}t)(1-\alpha)(A-\alpha)/2\right]}{(1-A)(1-\alpha)}\nonumber \\
 &  & +p\left[\mathcal{O}(\Delta x)+\mathcal{O}(\Delta\alpha A)\right]\nonumber \\
 & = & \frac{p}{1-A}\left[e^{-it}-e^{-it\lambda_{+}}\cos(\lambda_{-}t)+ite^{-it\lambda_{+}}f(\lambda_{-}t)\frac{A-\alpha}{2}\right]\nonumber \\
 &  & +p\left[\mathcal{O}(\Delta x)+\mathcal{O}(\Delta\alpha A)\right].\label{eq:0206-1}
\end{eqnarray}
 Since $\cos(\lambda_{-}t)$ and $f(\lambda_{-}t)$ only depend on
$\lambda_{-}^{2}=(\alpha-A)^{2}/4-\alpha A(1-u_{1}^{2})$ we have
\[
\cos(\lambda_{-}t)-itf(\lambda_{-}t)\frac{A-\alpha}{2}=\cos(\frac{\alpha-A}{2}t)-itf(\frac{\alpha-A}{2}t)\frac{A-\alpha}{2}+\mathcal{O}(\alpha A\Delta^{2}).
\]
So finally we get 
\begin{equation}
p\textrm{ term}=\frac{p}{1-A}\left[e^{-it}-e^{-itA}\right]+\mathcal{O}(s_{13}^{3}A\Delta)+\mathcal{O}(s_{13}\alpha A\Delta^{2}).\label{eq:0123-3}
\end{equation}

\subsection{Calculate $P^{(B)}-P^{(A)}$}

After expanding the square in $P^{(A)}$, we see the square term of
$p$ equals to the leading term in $P^{(B)}$. As for the square term
of $q$, since we have 
\begin{equation}
q=s_{12}c_{13}c_{12}c_{23}+\mathcal{O}(s_{13}),\label{eq:0206-2}
\end{equation}
so the differences of the corresponding term in $P^{(B)}$ and the
$q$ square term is 
\begin{eqnarray}
 &  & 4\alpha^{2}s_{12}^{2}c_{12}^{2}c_{23}^{2}\frac{\sin^{2}(A\Delta)}{A^{2}}-(q\textrm{ square term})\nonumber \\
 & = & 4\alpha^{2}\left[s_{12}^{2}c_{12}^{2}c_{23}^{2}(\frac{\sin^{2}(A\Delta)}{A^{2}}-\frac{\sin^{2}(\bar{A}\Delta)}{\bar{A}^{2}})+\mathcal{O}(s_{13}\Delta^{2})\right]\nonumber \\
 & = & 4\alpha^{2}\left[2\Delta^{2}\frac{1}{6}(\bar{A}^{2}-A^{2})+\mathcal{O}(s_{13}\Delta^{2})\right]\nonumber \\
 & = & \mathcal{O}(\alpha^{4}\Delta^{4})+\mathcal{O}(A\alpha^{3}\Delta^{4})+\mathcal{O}(\alpha^{2}s_{13}\Delta^{2})),\label{eq:0206-3}
\end{eqnarray}
where we have used the following approximation for $f(x)\equiv\frac{1}{x}\sin x$
\[
f(\bar{A}\Delta)-f(A\Delta)=-\frac{1}{6}(\bar{A}^{2}\Delta^{2}-A^{2}\Delta^{2})+\mathcal{O}(A^{4}\Delta^{4},\bar{A}^{4}\Delta^{4}).
\]

The remaining term is the cross term, computed as follows
\begin{eqnarray}
 &  & 2\textrm{Re}[\overline{p\textrm{ term}}\times q\textrm{ term}]\nonumber \\
 & = & 2\textrm{Re}[-2i\alpha\bar{p}q\frac{e^{2i\Delta}-e^{2iA\Delta}}{1-A}e^{-i(A+\alpha)\Delta}\frac{\sin(\bar{A}\Delta)}{\bar{A}}]\nonumber \\
 & = & 2\textrm{Re}[4\alpha(\frac{J_{CP}}{s_{\delta}}+\mathcal{O}(s_{13}^{2}))e^{i(\Delta-\alpha\Delta+\delta)}\frac{\sin[(1-A)\Delta]}{1-A}\frac{\sin(\bar{A}\Delta)}{\bar{A}}]\label{eq:0206-6}\\
 & = & 8\alpha[\frac{J_{CP}}{s_{\delta}}+\mathcal{O}(s_{13}^{2})][\cos(\Delta+\delta)+\mathcal{O}(\Delta\alpha)]\frac{\sin[(1-A)\Delta]}{1-A}[\frac{\sin(A\Delta)}{A}+\mathcal{O}(\alpha A\Delta^{3})+\mathcal{O}(\Delta^{3}\alpha^{2})]\nonumber \\
 & = & 8\alpha\frac{J_{CP}}{s_{\delta}}\cos(\Delta+\delta)\frac{\sin(A\Delta)}{A}\frac{\sin[(1-A)\Delta]}{1-A}\label{eq:0206-5}\\
 &  & +\mathcal{O}(\alpha s_{13}^{2}\Delta)+\mathcal{O}(s_{13}\alpha^{2}\Delta)+\frac{1}{6}s_{13}\mathcal{O}(\alpha^{3}\Delta^{3},\alpha^{2}A\Delta^{3}),
\end{eqnarray}
where $p=s_{13}c_{13}s_{23}e^{-i\delta}$ has been used. Combine the
result from eqs.(\ref{eq:0206-3}) and (\ref{eq:0206-5}), we have

\begin{equation}
P^{(B)}-P^{(A)}=\mathcal{O}(s_{13}^{2}\alpha\Delta)+\mathcal{O}(s_{13}\alpha^{2}\Delta^{2})+\mathcal{O}(\alpha^{3}A\Delta^{4})+\mathcal{O}(\alpha^{4}\Delta^{4}).\label{eq:0206-10}
\end{equation}

\begin{acknowledgments}
The author thanks E.Lisi, Hong-Jian He, Zhe Wang for early discussions
on the matter effect and the T2K experiment, and especially E. K.
Akhmedov and A.Yu. Smirnov for discussions on the main problem addressed
by this paper, also Werner Rodejohann for reading
the manuscript and useful suggestions, and Hiren Patel for improving the English writing of the manuscript. This work was supported by
the China Scholarship Council (CSC).
\end{acknowledgments}

\bibliographystyle{apsrev4-1}
\bibliography{ref}

\end{document}